\documentclass[aps,prd,preprintnumbers,showpacs,superscriptaddress,nofootinbib]{revtex4}
\usepackage{graphicx}
\usepackage{amsmath}
\advance\oddsidemargin by -\textwidth
\advance\oddsidemargin by -1in
\evensidemargin=\oddsidemargin
\def\fsl#1{\setbox0=\hbox{$#1$}           
   \dimen0=\wd0                                 
   \setbox1=\hbox{/} \dimen1=\wd1               
   \ifdim\dimen0>\dimen1                        
      \rlap{\hbox to \dimen0{\hfil/\hfil}}      
      #1                                        
   \else                                        
      \rlap{\hbox to \dimen1{\hfil$#1$\hfil}}   
      /                                         
   \fi}                                         %
\newcommand{\tr}{\mbox{tr}}
\newcommand{\VEV}[1]{\langle #1 \rangle}
\newcommand{\hyper}[4]{F\left(#1,#2,#3; #4\right)}

\newcommand{\arctanh}{\mbox{arctanh}}
\newcommand{\DxSB}{D$\chi$SB} 
\newcommand{\Q}{\Lambda_{\overline{\rm MS}}^{(D)}}

\begin{document}
\preprint{TU-645}
\preprint{DPNU-02-01}

\title{
    Dynamical chiral symmetry breaking \\ 
    in gauge theories with  extra dimensions
}
\author{Valery Gusynin}
\email[E-mail: ]{vgusynin@bitp.kiev.ua}
\affiliation{Bogolyubov Institute for Theoretical Physics, 03143, Kiev, Ukraine}
\affiliation{Department of Physics, Nagoya University, Nagoya 464-8602, Japan}
\author{Michio Hashimoto}
\email[E-mail: ]{michioh@tuhep.phys.tohoku.ac.jp, michioh@post.kek.jp}
\affiliation{Department of Physics, Tohoku University, Sendai 980-8578, Japan}
\affiliation{Theory Group, KEK, Oho 1-1, Tsukuba, Ibaraki 305-0801, Japan}
\author{Masaharu Tanabashi}
\email[E-mail: ]{tanabash@tuhep.phys.tohoku.ac.jp}
\affiliation{Department of Physics, Tohoku University, Sendai 980-8578, Japan}
\author{Koichi Yamawaki}
\email[E-mail: ]{yamawaki@eken.phys.nagoya-u.ac.jp}
\affiliation{Department of Physics, Nagoya University, Nagoya 464-8602, Japan}

\begin{abstract}
We investigate 
dynamical chiral symmetry breaking in vector-like
gauge theories in $D$ dimensions 
with ($D-4$) compactified extra dimensions, based on
the gap equation (Schwinger-Dyson equation) 
and the effective potential 
for the bulk gauge theories 
within the improved ladder
approximation. 
The 
non-local gauge fixing
method is adopted so as to keep the ladder 
approximation consistent with the Ward-Takahashi identities.

Using the one-loop
$\overline{\rm MS}$ gauge coupling of the truncated KK effective
theory which 
has a nontrivial ultraviolet fixed point (UV-FP)
$g_*$ for the (dimensionless) bulk gauge coupling ${\hat g}$,
we 
find 
that there exists a 
critical number of flavors,
$N_f^{\rm
  crit}$ ($\simeq 4.2, 1.8$  for  $D=6, 8$ for $SU(3)$ gauge theory):
For $N_f
>N_f^{\rm crit}$, 
the dynamical chiral symmetry breaking takes place not only in the
``strong-coupling phase'' (${\hat g} >g_*$) but also in the 
``weak-coupling phase'' (${\hat g} <g_*$) 
when the cutoff is large enough. 
 For $N_f
< N_f^{\rm crit}$, on the other hand,
only the strong-coupling phase is a broken phase
and we can formally define a continuum 
(infinite cutoff) limit, so that the physics is insensitive to the cutoff
in this case.

We also perform a similar analysis using the one-loop ``effective  gauge
coupling''. 
We find the 
$N_f^{\rm crit}$ 
turns out to be a value similar
to that of the $\overline{\rm MS}$ case,  notwithstanding 
the enhancement of the coupling compared with that of the $\overline{\rm MS}$.

\end{abstract}

\maketitle

\section{Introduction}
Much attention has been paid to the extra dimension physics,
particularly the large scale scenarios~\cite{ADD,antoni90}. 
Although the 
notion of the ``extra dimension'' might be ``deconstructed'' in terms
of certain renormalizable four-dimensional gauge theories~\cite{georgi01,hill01},
there still exist attractive features of gauge theories with extra dimensions
which deserve further non-perturbative studies.

As such three of us~\cite{TMSM_ED} (referred to as (I) hereafter)
have recently studied dynamical symmetry breaking 
of the top-mode standard model (TMSM)~\cite{MTY89,Nambu89,BHL} 
in a version in $D$ dimensions, with the $D-4$ being compactified extra 
dimensions (ACDH scenario)~\cite{ACDH}. 
The ACDH scenario
was based on earlier proposal \cite{TC_ED} of the TMSM with extra
dimensions, 
which was motivated by the topcolor ideas~\cite{kn:topcolor}, 
and found 
that the (dimensionless)  bulk QCD coupling above 
the compactified scale becomes 
strong due to KK modes contributions 
and hence may trigger the top quark condensate without ad hoc four-fermion
interactions as in the original TMSM\@.\footnote{
It is discussed in Ref.~\cite{kn:kobakhidze} to combine the original
version of TMSM with the space-time having extra dimensions.  
}
Unlike Ref.~\cite{TC_ED} in which only $t_R$ is in the bulk, somewhat
simpler situation is assumed in the ACDH scenario: 
the entire third family lives in the bulk, which enables us to
investigate dynamically whether the top condensate really takes
place in this scenario.
In (I) we used the $D$-dimensional gap equation (improved ladder
Schwinger-Dyson (SD) equation \cite{imp_ladder}) with one-loop
$\overline{\rm MS}$ 
running (bulk) gauge coupling of the truncated KK effective
theory \cite{DDG} for such a purpose.\footnote{
Dynamical chiral symmetry breaking in other types of models with
extra dimensions is studied in Ref.~\cite{kn:hiroshima}.
}
In (I) we found~\cite{TMSM_ED}
that the (dimensionless) bulk  gauge 
coupling $\hat g$ has a nontrivial ultraviolet fixed point (UV-FP) $g_*$
in the same one-loop truncated KK effective theory as that ACDH was based on. 
Since  the running bulk  gauge coupling rapidly reaches
the UV-FP, the gap equation is essentially governed by 
the UV-FP and can well be approximated by that with the running coupling
replaced by the UV-FP value: ${\hat g}^2 \rightarrow g_*^2$
(``gap equation on the UV-FP''). 
If we assume the UV-FP persists non-perturbatively, then 
the bulk QCD coupling is in the weak coupling region ($\hat g <g_*$), 
since the coupling to be matched with the 3-brane QCD coupling at the
compactification scale is certainly a weak coupling there and hence
never exceeds the UV-FP\@.
This implies that top quark condensate is only possible when 
 $g_* >\hat g_{\rm crit}$ ($\kappa_D >\kappa_D^{\rm crit}$ in the notation
 of (I), with $\kappa_D$ proportional to $g_*^2$), where  $\hat g_{\rm crit}$  is
determined by the SD gap equation just on the UV-FP mentioned above.
We then found that top quark condensate cannot take place
in the simplest case of the ACDH scenario, 
$D=6$ and $N_f=2$ (only the third family besides
gauge bosons lives in the bulk), where we found 
the UV-FP value is smaller than the 
critical value ($g_* < \hat g_{\rm crit}$), 
while in the case with $D=8$ and $N_f=2$ we found it can 
($g_* > \hat g_{\rm crit}$).

In (I) we further studied the phase structure of gauge theories
 in $D$ dimensions with the $D-4$ dimensions compactified,
 not restricted to the TMSM\@.
Since the gap equation on the UV-FP possesses a  
scale invariance, the phase transition takes the form of
``conformal phase transition'' \cite{MY97} having an 
essential-singularity-type scaling. 
When $g_* > \hat g_{\rm crit}$, 
the dynamical mass
function has a slowly damping asymptotic behavior which corresponds to
a large anomalous dimension $\gamma_m =D/2-1$, somewhat similar to the
walking technicolor \cite{walking}.

We also discussed in (I) another gap equation, with the 
one-loop $\overline{\rm MS}$ running coupling replaced by the
one-loop ``effective coupling'' which includes finite
renormalization effects. Since the effective coupling turned out to
be considerably enhanced compared with the  $\overline{\rm MS}$ one,
we argued that there might exist a possibility that 
even the simplest case of ACDH scenario
with $D=6$ and $N_f=2$ may give rise to a top quark condensate.
 
In this paper, we further study the non-perturbative dynamics of 
various vector-like gauge theories with extra dimensions, not restricted to
the TMSM\@. Since $g_*^2$ (or $\kappa_D$) is written in terms of $N_f$,
we find that there exists a {\it critical number of flavors} $N_f^{\rm crit}$, 
such that $N_f >N_f^{\rm crit}$ for $g_*> \hat g_{\rm crit}$.
 We find that there exists a rich phase structure in such theories:
The phase is separated not only into  $N_f >N_f^{\rm crit}$ 
($g_*> \hat g_{\rm crit}$) and $N_f <N_f^{\rm crit}$ 
($g_*< \hat g_{\rm crit}$), but also into ${\hat g}> g_*$ (strong coupling phase)
and ${\hat g}< g_*$ (weak coupling phase) (See Fig.\ref{fig:phase_diagram}). 
This may be useful for a large variety
of model buildings beyond the standard model.

In order to 
systematically study the SD gap equation
in a manner consistent with the Ward-Takahashi identity,
we adopt 
a so-called 
{\it 
non-local gauge fixing}. 
Actually, as is known in the
four-dimensional case~\cite{KM_chiWT}, the chiral Ward-Takahashi identity
is violated in 
the gap equation of (I) which is ``improved'' from
the ladder SD equation by a simple ansatz to replace the constant
(dimensionful) bulk gauge coupling $g_D$ by the running one as~\cite{imp_ladder}:
\begin{equation}
  g_D^2 \rightarrow g_D^2(\max(-p^2,-q^2)),
\end{equation}
where $p^\mu$ and $q^\mu$ are the momenta of external and loop fermions,
respectively. This problem can be 
solved by taking the running coupling as~\cite{KM_chiWT} 
\begin{equation}
g_D^2 \rightarrow g_D^2(-(p-q)^2),
\end{equation}
namely a function of gauge boson loop momentum.
Then the Landau gauge used in (I) no longer guarantees 
$A(-p^2)\equiv 1$, 
which is then inconsistent with 
the bare vertex ansatz of the ladder approximation.
This problem can also be 
remedied by employing the so-called non-local gauge 
fixing~\cite{NLG,Simmons90,KM_chiWT,NLG_kondo},
by which the gauge parameter is arranged to be momentum-dependent 
so as to keep $A(-p^2)\equiv 1$.
Note that the above problems are numerically not serious in the 
four-dimensional cases and the method of (I) is widely used accordingly.
However, situation in the higher dimensional case with power running coupling
may drastically  be changed. 

We first re-analyze the gap equation on the UV-FP with ${\hat g} \equiv g_*$,
in which the dynamical symmetry breaking takes place for 
$g_* > \hat g_{\rm crit}$ ($N_f > N_f^{\rm crit}$) as in (I):
As a result of the above more sophisticated treatment, however,
we find that $\hat g_{\rm crit}^2$ is larger than
that of (I) by a factor $D/4$, which is a substantial change 
for $D \gg 4$. This result implies that the dynamical symmetry
breaking gets suppressed compared with the result of (I).
For the $SU(3)$ gauge theory our new gap
equation yields:
\begin{equation}
  N_f^{\rm crit} \simeq 
  \begin{cases}
    4.2,  & \text{for $D=6$}, \\
    1.8,  & \text{for $D=8$}, \\
    0.8,  & \text{for $D=10$}. \\
  \end{cases}
  \label{eq:Nfcrit}
\end{equation}
   
Based on the  gap equation both {\it on and off the UV-FP},
we further reveal a {\it full phase structure} in the 
${\hat g}^2$-$N_f$ plane: Although
the solution of the gap equation {\it on the UV-FP}, as in the
analysis of (I), 
just separates the phases by  $N_f >N_f^{\rm crit}$ ($g_* > \hat g_{\rm crit}$)
and  $N_f <N_f^{\rm crit}$ ($g_* < \hat g_{\rm crit}$), we here also analyze
the gap equation {\it off the UV-FP} which further separates the phases by 
${\hat g} > g_*$ (``strong-coupling phase'') and
${\hat g} < g_*$ (``weak-coupling phase'').
 
 For $N_f>N_f^{\rm crit}$ ($g_* >\hat g_{\rm crit}$), 
 which is the case we studied in (I),
the dynamical chiral symmetry breaking takes place not only in the
strong-coupling phase (${\hat g}> g_*$) but also in the weak-coupling
 one (${\hat g}< g_*$)  
as far as the cutoff $\Lambda$ is large enough (namely, 
${\hat g}(\mu=\Lambda)$ is rather close to $g_*$).   
This case is relevant to the TMSM with extra dimensions
(ACDH scenario) \cite{ACDH} whose bulk QCD 
coupling is matched with that of the brane QCD at the compactification scale 
which is certainly weak and hence the theory necessarily should be set 
in the weak-coupling phase.
In order to have dynamical symmetry breaking even in the weak-coupling phase,
we need to arrange $N_f> N_f^{\rm crit}$ : 
{}From the result Eq.(\ref{eq:Nfcrit})
we conclude that the simplest version of the ACDH scenario with $N_f=2$  
 does not give rise to a top quark condensate for $D=6$,
while it can for $D=8$ and $D=10$.

 For $N_f > N_f^{\rm crit}$, we further find it impossible
to define the continuum limit, despite the fact that  
the essential-singularity type scaling with respect to $\kappa_D$
found in (I) superficially suggests a conformal phase transition
having a large anomalous dimension $\gamma_m =D/2-1$:
Actually the value of $\kappa_D$ is not continuous and hence cannot be
taken arbitrarily close to $\kappa_D^{\rm crit}$. 
Then the UV cutoff should be considered as a physical one and the
low-energy physics remains cutoff-sensitive in this case.

Moreover, in the ACDH scenario the scale of the physical UV cutoff $\Lambda$
{\it is no longer an adjustable parameter but a ``predictable one''}
in contrast to the treatment in Ref.~\cite{ACDH}, since the bulk gauge coupling
${\hat g}$ is completely controlled by the 3-brane QCD coupling at the
compactification scale 
and the KK effective theory, and hence $\Lambda$ is uniquely tied up with the
dynamical mass of the condensed fermion (top quark) through the gap equation.
If we use as an input the value of $F_\pi(\simeq 250 {\rm GeV})$ which is
also tied up 
with the top quark mass, then the cutoff is ``predicted'' in terms of
$F_\pi$. 
The situation is completely different from the original TMSM 
where $\Lambda$ is related through 
the gap equation only to two parameters:
the dynamical mass {\it and} the four-fermion coupling which 
is a free parameter.  This implies that even when the model
is arranged as $N_f> N_f^{\rm crit}$, the phenomenological analysis of 
Ref.~\cite{ACDH} should be modified substantially by taking account of
this fact, 
which we shall report in  a separate paper. 

For $N_f< N_f^{\rm crit}$ ($g_*< \hat g_{\rm crit}$), 
on the other hand, we find a {\it novel
situation}: The
strong-coupling phase (${\hat g} > g_*$) is in 
the chiral-symmetry broken phase, while 
the weak-coupling phase (${\hat g}<g_*$) is in the unbroken one,
and we can formally define a continuum limit
(infinite cutoff limit) at the phase boundary ${\hat g} =g_*$
 {\it with a large anomalous
dimension} $\gamma_m = (D/2-1) (1-\tilde \nu)$ 
($0<\tilde \nu \equiv \sqrt{1-(g_*/\hat g_{\rm crit})^2}<1$), 
a situation similar to the ladder QED \cite{Miransky}.
This fact implies that the low-energy physics becomes {\it insensitive to
the details of 
the physics around cutoff} (stringy physics ?). 
Then, no matter
whatever physics may exist at the cutoff,
the strong-coupling phase of this case may be useful within the
framework of local field theory (without referring to, e.g., stringy
physics) for model building, such as a
 ``bulk technicolor''.  The bulk technicolor then resembles the walking 
technicolor~\cite{walking} 
with large anomalous dimension and hence is expected to be free 
{}from FCNC problems.

We also perform a similar analysis using the one-loop ``effective gauge
coupling''. 
Although the gauge boson propagator explicitly depends on the UV
cutoff and cannot be renormalized in this scheme, we find a critical
$N_f$ 
turns out to be a value similar 
to that of the $\overline{\rm MS}$ case, 
$4< N_f^{\rm crit} <5$ $(D=6)$. This is rather surprising,
 considering the fact that as we showed in (I) the effective coupling is
enhanced  roughly double
compared with that of the $\overline{\rm MS}$.
Then this result strongly suggests that for all ambiguities  of
the approximations of the gap equation,
the simplest case of ACDH scenario with $N_f=2$ is quite unlikely for $D=6$.
 
The paper is organized as follows:
In Section 2 we write down the SD gap equation in $D$ dimensions with the
non-local gauge fixing. In Section 3 we obtain analytical solution 
to the SD equation on the UV-FP with the running coupling
set to be just on the UV-FP, ${\hat g} \equiv g_*$. 
The ground state is identified through the Cornwall-Jackiw-Tomboulis (CJT) 
effective potential \cite{kn:CJT}. 
In Section 4 we present a full
phase structure in ${\hat g}^2$-$N_f$ plane,
based on the solution of the gap equation both on and off the UV-FP:
For $N_f >N_f^{\rm crit}$ we find that both the strong-coupling 
phase (${\hat g}>g_*$) and the weak coupling phase (${\hat g} <g_*$) 
are broken phases and is relevant to the ACDH scenario of
the TMSM whose model building is then constrained by the value of
$N_f^{\rm crit}$. 
We also find no continuum limit in this case
and the cutoff is predictable in terms of $F_\pi\simeq 250 {\rm GeV}$.
On the other hand, for $N_f <N_f^{\rm crit}$ we find that the UV-FP separates  
a broken phase (for strong-coupling phase) and an unbroken phase 
(for the weak coupling phase):
We can formally define a continuum limit at the UV-FP with
large anomalous dimension and the theory may be useful
for ``bulk technicolor''. In Section 5 we analyze the gap equation
with the effective coupling instead of
the  $\overline{\rm MS}$ running coupling through the non-local
gauge fixing. 
We also find a mean-field scaling. Section 6 is for Summary and Discussions.
Appendix~\ref{app:eff_pot} is devoted to yet another effective
potential than the CJT potential,
which has a more direct relevance to the bound states picture.
In Appendix~\ref{schro} we present a Schr\"{o}dinger-like equation
which yields some  
intuition on the $D$-dependence of the phase transition.
Effects of the infrared cutoff in the gap equation are discussed in
Appendix~\ref{ir-cutoff}. 
Appendix~\ref{eff_coupl} gathers formulas of effective coupling.

\section{Gap equation with the non-local gauge fixing}

Although the $D$-dimensional Lorentz symmetry is explicitly violated
by the compactification of the extra dimensions, such a effect should
be proportional to the inverse of the compactification radius $R^{-1}$.
For sufficiently large momentum $|p^2| \gg R^{-2}$, we thus expect
that the $D$-dimensional Lorentz symmetry is restored approximately,
which enables us to make an ansatz for the bulk fermion propagator in a
Lorentz covariant form:
\begin{equation}
  iS^{-1}(p)=A(-p^2)\fsl{p} - B(-p^2).
\end{equation}

The appearance of the non-zero fermion mass (gap) $B(-p^2) \ne 0$ is a
signal of the chiral symmetry breaking in the bulk.
The aim of this section is to construct an appropriate gap equation,
by which we investigate the chiral phase transition in the
vector-like gauge theories with the extra dimensions. 

Let us start with the naive ladder approximation of the
Schwinger-Dyson equation of the bulk fermion propagator \cite{KN89}
\begin{widetext}
\begin{eqnarray}
  A(-p^2) &=& 
    1 + \dfrac{C_F}{-p^2}\int \!\! \dfrac{d^D q}{(2\pi)^D i }
        \dfrac{A(-q^2)}{-A^2(-q^2) q^2 +B^2(-q^2)} 
    \nonumber\\
    & & \qquad \times 
        \left[
          -(3-D-\xi) \dfrac{p \cdot q}{(p-q)^2}
          +2(1-\xi) \dfrac{p\cdot (p-q) q\cdot (p-q)}{(p-q)^4}
        \right]g_D^2 ,
\label{eq:sdA}
   \\
  B(-p^2) &=&
    m_0 + C_F \int\!\! \dfrac{d^D q}{(2\pi)^D i }
        \dfrac{B(-q^2)}{-A^2(-q^2) q^2 +B^2(-q^2)} \cdot
        \dfrac{(D-1+\xi) g_D^2}{-(p-q)^2},
\label{eq:sdB}
\end{eqnarray}
\end{widetext}
with $C_F$ being the Casimir of the fermion representation 
($C_F=(N^2-1)/(2N)$ for the fundamental representation of $SU(N)$
gauge group). 
Here
$g_D$, $\xi$ and $m_0$ are the bulk gauge coupling strength, the gauge
fixing parameter and the fermion bare mass, respectively. 
It should be noted that the mass dimension of the gauge coupling
strength $g_D^2$ is negative, $-\delta$, for $D=4+\delta > 4$.

Within the naive ladder approximation, the effect of the running gauge
coupling strength is completely ignored, however.
In order to remedy this drawback, the gauge coupling constant $g_D$
needs to be replaced by something in which the running effect is
incorporated appropriately. 

Since there exist 3 different momenta, $x\equiv -p^2$, $y\equiv
-q^2$, $z\equiv -(p-q)^2$, in the Schwinger-Dyson Eqs.(\ref{eq:sdA})
and (\ref{eq:sdB}),  
there exist various ways to incorporate the running effect in the
gap equation.
In Ref.~\cite{TMSM_ED}, we improved the Schwinger-Dyson equation using a simple
replacement \cite{imp_ladder}, 
\begin{equation}
  g_D^2 \rightarrow g_D^2(\max(x,y)),
\label{eq:imp1}
\end{equation}
with $g_D^2(\max(x,y))$ being the running gauge coupling.
This prescription is widely used and has various technical advantages:
The angular integrals in the gap equation can be performed in an
analytical manner. 
The wave function factor $A$ is shown to be unity in the Landau gauge
$\xi=0$, which makes the ladder approximation consistent with the 
vector Ward-Takahashi identity.

Although the prescription Eq.(\ref{eq:imp1}) has been used widely in
the analysis of the dynamical chiral symmetry breaking in the four
dimensional QCD\@, 
it is pointed out \cite{KM_chiWT} that Eq.(\ref{eq:imp1}) is not
consistent with the 
chiral Ward-Takahashi identity, if the same prescription is applied to
the axial-vector vertex.
 
In this paper, we therefore use a different choice \cite{KM_chiWT}
\begin{equation}
  g_D^2 \rightarrow g_D^2(z),
\label{eq:imp2}
\end{equation}
in which the gauge boson momentum $z$ is used as the scale of the
running gauge coupling strength.
The prescription Eq.(\ref{eq:imp2}) is consistent with the
chiral Ward-Takahashi identity, but it induces non-trivial wave
function factor $A$ within conventional gauge fixing methods,
leading to an inconsistency with the ladder approximation and the
vector Ward-Takahashi identity.
   
In order to avoid such a dilemma, we use the non-local gauge fixing
method.
The method was originally invented in the analysis of four dimensional
gauge theories \cite{NLG} and extended to gauge theories in $D$
dimensions \cite{Simmons90}. 
It was then reformulated into a compact formula in the
analysis of four dimensional QCD by using a different
approach \cite{KM_chiWT}.
The method of Ref.~\cite{KM_chiWT} is extended to gauge theories in
arbitrary dimensions \cite{NLG_kondo}.
Here we give a brief derivation of the non-local gauge in order to
explain notations used in this paper.

The non-local gauge fixing method is based on the observation that the
parameter $\xi$ can be generalized to a function of the momentum $z$,
$\xi(z)$ by introducing the non-local gauge fixing operator.
It is then possible to choose the specific form of $\xi(z)$ so as to
make the wave function factor $A\equiv1$.

We start with the Schwinger-Dyson equation of the fermion wave
function $A$, Eq.(\ref{eq:sdA}).
After the Wick rotation, it reads
\begin{widetext}
\begin{equation}
  A(x) = 
    1 + \dfrac{C_F}{x} \int_0^{\Lambda^2} dy y^{D/2-1}
        \dfrac{A(y)}{A^2(y) y +B^2(y)} K_A(x,y). 
\end{equation}
We introduced the ultraviolet cutoff $\Lambda$, where the
$D$-dimensional effective field theory is considered to be replaced by 
yet unknown underlying physics (e.g., string theory, (de)constructed
extra dimensions \cite{georgi01,hill01}, etc.).

The integral kernel $K_A$ is given by
\begin{eqnarray}
  K_A(x,y) 
  &=&
    \dfrac{\Omega_{\rm NDA}}{B(\frac{1}{2},\frac{D}{2}-\frac{1}{2})}
    \int_0^\pi d\theta (\sin\theta)^{D-2}
    g_D^2(z) 
  \nonumber\\
  & & \qquad \qquad \qquad \times 
    \left[
      (D-1-\xi(z))\dfrac{\sqrt{xy}\cos\theta}{z}
     -2(1-\xi(z)) \dfrac{xy}{z^2} \sin^2\theta
    \right],
\label{eq:kernA}
\end{eqnarray}
with $\Omega_{\rm NDA}$ being the naive dimensional analysis \cite{kn:NDA} (NDA)
factor 
\begin{equation}
  \Omega_{\rm NDA} \equiv \dfrac{1}{(4\pi)^{D/2}\Gamma(D/2)}.
\end{equation}
The angle $\theta$ is the angle between Euclidean momenta $q_E$ and
$p_E$, 
\begin{equation}
  z = x + y - 2\sqrt{xy} \cos\theta.
\label{eq:wdef}
\end{equation}
Noting 
\begin{displaymath}
  (\sin\theta)^{D-2} \cos\theta 
    = \dfrac{1}{D-1} \dfrac{d}{d\theta}\left((\sin\theta)^{D-1}\right),
\end{displaymath}
we find 
\begin{eqnarray}
  K_A(x,y)
  &=&
    \dfrac{\Omega_{\rm NDA}}{B(\frac{1}{2},\frac{D}{2}-\frac{1}{2})}
    \Biggl\{
      -\dfrac{1}{D-1}\int_0^\pi d\theta (\sin\theta)^{D-1}\sqrt{xy} 
       \dfrac{d}{d\theta}\left((D-1-\xi(z))g_D^2(z)\dfrac{1}{z}\right)
  \nonumber\\
  & & -2\int_0^\pi d\theta (\sin\theta)^D (1-\xi(z))g_D^2(z)\dfrac{xy}{z^2}
    \Biggr\},
\label{eq:kernA2}
\end{eqnarray}
where we have integrated $d/d\theta$ by parts.

The $\theta$ differentiation  $d/d\theta$ in Eq.(\ref{eq:kernA2}) can
be written as 
\begin{displaymath}
  \dfrac{d}{d\theta}
  = \dfrac{dz}{d\theta} \dfrac{d}{dz}
  = 2\sqrt{xy} \sin\theta \dfrac{d}{dz}.
\end{displaymath}
We then obtain
\begin{eqnarray}
  K_A(x,y) 
  &=&
    -2\dfrac{\Omega_{\rm NDA}}{B(\frac{1}{2},\frac{D}{2}-\frac{1}{2})}
    xy \int_0^\pi d\theta (\sin\theta)^D \nonumber\\
  & & \qquad \qquad \times
    \left[
      \dfrac{1}{D-1}
      \dfrac{d}{dz}\left((D-1-\xi(z))\dfrac{g_D^2(z)}{z}\right)
     +(1-\xi(z))\dfrac{g_D^2(z)}{z^2}
    \right].
\end{eqnarray}

The condition $K_A\equiv 0$ can be guaranteed if $\xi(z)$ satisfies a
differential equation 
\begin{equation}
 \dfrac{1}{D-1}
   \dfrac{d}{dz}\left((D-1-\xi(z))\dfrac{g_D^2(z)}{z}\right)
 +(1-\xi(z))\dfrac{g_D^2(z)}{z^2}
  = 0.
\label{eq:non-local}
\end{equation}
It is easy to solve Eq.(\ref{eq:non-local}). 
We find the solution (non-local gauge) is given by
\begin{equation}
  \xi(z) =
    \dfrac{D-1}{g_D^2(z) z^{D-2}} 
    \int_0^z dz z^{D-2} \dfrac{d}{dz} g_D^2(z),
\label{eq:non-local2}
\end{equation}
where the integration constant is taken so as to make $\xi(z)$ regular 
at $z=0$.

Using the non-local gauge fixing parameter Eq.(\ref{eq:non-local2}),
the wave function factor $A$ can be set unity.
The gap equation (\ref{eq:sdB}) then reads,
\begin{equation}
  B(x) =
    m_0 + C_F \int_0^{\Lambda^2} dy y^{D/2-1}
        \dfrac{B(y)}{y +B^2(y)} K_B(x,y),
\label{eq:gapeq1}
\end{equation}
where
\begin{equation}
  K_B(x,y) \equiv
    \dfrac{\Omega_{\rm NDA}}{B(\frac{1}{2},\frac{D}{2}-\frac{1}{2})}
    \int_0^\pi d\theta (\sin\theta)^{D-2}
    \dfrac{(D-1+\xi(z))g_D^2(z)}{z}.
\label{eq:kernB}
\end{equation}
Eqs.(\ref{eq:non-local2}), (\ref{eq:gapeq1}) and (\ref{eq:kernB}) are
our basic equations to be solved in this paper.

\end{widetext} 

It should be kept in mind that the gap equation Eq.(\ref{eq:gapeq1})
is not valid for  $x, y \lesssim R^{-2}$ ($x, y \equiv
|p^2|, |q^2|$) due to the compactification of the extra dimensions.
In order to estimate uncertainties coming from the compactification
sensitive infrared region, 
we introduce an infrared cutoff $M_0^2 \sim R^{-2}$ on and off in the
following analyses. 
We will actually find that many results are insensitive to $M_0^2$ if
the ultraviolet cutoff is taken to be large sufficiently 
$\Lambda^2 \gg M_0^2$.

\section{Solution at the fixed point}
\label{sec:fp_solution}

We next consider the running of the gauge coupling in theories with
extra dimensions compactified to an orbifold $T^\delta/Z_n$ with
radius $R$.
Here $Z_n$ represents the discrete group with order of $n$. 

In Ref.~\cite{TMSM_ED}, the {\em dimensionless} bulk gauge coupling
$\hat g$ is defined as,
\begin{equation}
  \hat g^2 \equiv \dfrac{(2\pi R\mu)^\delta}{n} g^2, \label{eq:dimless} 
\end{equation}
with $g$ being the gauge coupling of the truncated KK effective
theory \cite{DDG}. 
The bulk gauge coupling $g_D$ is given by
\begin{equation}
  g_D^2 = \dfrac{\hat g^2(\mu)}{\mu^\delta}.
\end{equation}
It is then shown that the {\em dimensionless} bulk gauge
coupling obeys the renormalization group equation (RGE),
\begin{equation}
  \mu \frac{d}{d\mu} \hat g = \frac{\delta}{2} \hat g
  + (1+\delta/2) \Omega_{\rm NDA} b' \hat g^3,
\label{eq:rge1}
\end{equation}
at the one-loop approximation of the $\overline{\rm MS}$ coupling of
the truncated KK effective theory.
The RGE factor $b'$ is given by
\begin{equation}
  b' = -\frac{26-D}{6} C_G + \frac{\eta}{3} T_R N_f,
\label{eq:betafun}
\end{equation}
where $\eta$ represents the dimension of the spinor representation of
$SO(1,D-1)$,
\begin{equation}
  \eta \equiv \tr_\Gamma 1 = 2^{D/2}\qquad
  \mbox{for even $D$},
\end{equation}
and $N_f$ is the number of fermions in the bulk.
The group-theoretical factor $C_G$ and $T_R$ are given by $C_G=N$ and
$T_R=1/2$ for $SU(N)$ gauge theory.

It is interesting to note that the {\em dimensionless} gauge coupling
$\hat g$ has a non-trivial asymptotically stable ultraviolet fixed
point (UV-FP),
\begin{equation}
  g_*^2 \Omega_{\rm NDA} =
  \dfrac{1}{-\left(\frac{2}{\delta}+1\right)b'},
\label{eq:uvfp1}
\end{equation}
for $b'<0$ or 
\begin{equation}
  N_f < N_f^{\rm ANS} \equiv \dfrac{(26-D)C_G}{2\eta T_R}.
\end{equation}

On the other hand, 
the coupling $\hat g$ grows without a bound in the high energy
region (asymptotically-not-stable) and the UV-FP disappears
for $b' > 0$ or $N_f > N_f^{\rm ANS}$.
Hereafter we thus restrict ourselves to the analysis of gauge theories with
$N_f < N_f^{\rm ANS}$.

\begin{figure}[htbp]
  \begin{center}
    \resizebox{0.45\textwidth}{!}{
      \includegraphics{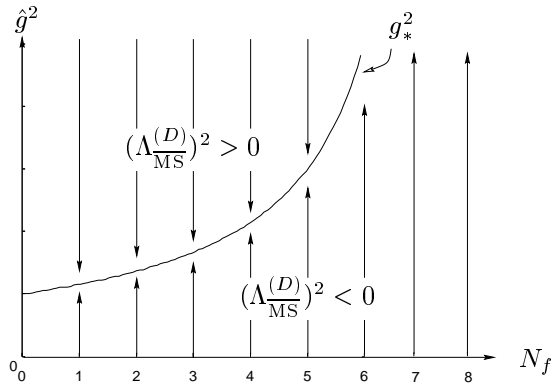}}
    \caption{
      The one-loop renormalization group flow of $\hat g^2$ in
      $SU(3)$ gauge theory in $D=4+2$ dimensions for various $N_f$.
      The gauge coupling $\hat g$ approaches its UV-FP for $\mu
      \rightarrow \infty$.
}
    \label{fig:rgeflow}
  \end{center}
\end{figure}

It is straightforward to solve RGE Eq.(\ref{eq:rge1}).
Particularly,
the coupling $\hat g^2$ behaves as 
\begin{equation}
  \hat g^2(\mu^2) = \dfrac{\mu^2 g_*^2}{\mu^2-(\Q)^2},
\label{eq:msbar6}
\end{equation}
in $D=4+2$ dimensional gauge theories.
Here $(\Q)^2$ is the scale parameter of the theory.
Vanishing of $(\Q)^2=0$ corresponds to the UV-FP solution ($\hat g^2 =
g_*^2$) and it
implies that the theory becomes approximately scale-invariant, except that
the scale-invariance is violated by the cutoff $\Lambda$ and the
compactification scale $R^{-1}$. 
On the other hand, positive $(\Q)^2>0$ (negative $(\Q)^2<0$) corresponds to
strongly interacting phase $\hat g^2 > g_*^2$ (weakly interacting
phase  $\hat g^2 < g_*^2$). See Figure \ref{fig:rgeflow}. 

Although the renormalization group structure calculated above in the
one-loop level cannot be justified within a perturbative analysis, 
the existence of the non-trivial UV-FP is  supported by
a recent lattice calculation in a certain case. \cite{EKM} 
Moreover, the existence of such a UV-FP may open an 
interesting possibility in the model building with the compactified
extra dimensions. 
Absence of such a non-trivial UV-FP is implicitly assumed in the
conventional models with extra dimensions, in which physical UV cutoff needs
to be introduced and the predictions depend on the
non-field-theoretical physics at the cutoff scale (e.g., stringy
physics). 
On the other hand, 
if there exists a non-trivial UV-FP in the model with extra dimensions, 
the low energy physics can be predicted almost entirely from the field 
theoretical properties of the UV-FP\@.
The low energy predictions become insensitive to the physics at the
UV cutoff.

We believe that this new possibility is interesting enough to justify
the investigation of the dynamical properties of the chiral phase
transition around the presumed UV-FP\@.
It should be noted, however, the
value of the one-loop UV-FP Eq.(\ref{eq:uvfp1}) can be
affected substantially by the higher-loop or non-perturbative
effects. 
Nevertheless, we adopt the one-loop formulas Eqs.(\ref{eq:rge1}) and
(\ref{eq:uvfp1}) in the following sections,
assuming optimistically that qualitative behavior can be obtained
within one-loop formulas.

We are now ready to start the analysis of the gap equation. 
The running effect of the gauge coupling is taken into account by
replacing the renormalization scale ($\mu$) dependence with the gauge
boson momentum $z$: 
\begin{equation}
  g_D^2(z) = \dfrac{\hat g^2 (\mu=\sqrt{z})}{z^{\delta/2}}.
\end{equation}

We start with the simplest case where the dimensionless
gauge coupling is standing at the UV-FP $g_*^2$:
\begin{equation}
  g_D^2(z) = \dfrac{g_*^2}{z^{\delta/2}}.
\label{eq:q=0}
\end{equation}
Plugging Eq.(\ref{eq:q=0}) into Eq.(\ref{eq:non-local2}) we find that
the $A\equiv 1$ gauge is given by a simple form
\begin{equation}
  \xi(z) = - \dfrac{(D-1)(D-4)}{D}.
\label{eq:nlg1}
\end{equation}
We note that Eq.(\ref{eq:nlg1}) is merely a constant.
The gauge fixing operator is thus a local one in this case at the
UV-FP\@.

It is straightforward to perform the angular integral in
Eq.(\ref{eq:kernB}),
\begin{widetext} 
\begin{equation}
  K_B(x,y) = \dfrac{4(D-1)}{D} g_*^2 \Omega_{\rm NDA}
  \left(
    \dfrac{1}{x^{D/2-1}}\theta(x-y) +\dfrac{1}{y^{D/2-1}}\theta(y-x)
  \right).
\end{equation}
We thus obtain the gap equation
\begin{equation}
  B(x) = m_0 + \dfrac{4(D-1)}{D} \kappa_D 
    \int_0^{\Lambda^2}
      dy \dfrac{y^{D/2-1} B(y)}{y + B^2(y)}\left(
    \dfrac{1}{x^{D/2-1}}\theta(x-y) 
   +\dfrac{1}{y^{D/2-1}}\theta(y-x) 
  \right),
\label{eq:gap2}
\end{equation}
with
\begin{equation}
  \kappa_D \equiv C_F g_*^2 \Omega_{\rm NDA}.
\end{equation}
We next try to solve the gap equation (\ref{eq:gap2}) analytically. 

Differentiating Eq.(\ref{eq:gap2}) over $x$, we find that the integral 
equation (\ref{eq:gap2}) is equivalent to the differential equation
\begin{equation}
  \frac{d}{dx}\left[ x^{D/2} \frac{d}{dx} B \right]
  + \dfrac{2(D-1)(D-2)}{D} \kappa_D \dfrac{x^{D/2-1} B}{x+B^2} = 0,
\label{eq:diff1}
\end{equation}
and a set (infrared and ultraviolet) of boundary conditions:
\begin{eqnarray}
  \left. x^{D/2} \frac{d}{dx} B(x) \right|_{x=0} = 0, \qquad
  & & (\mbox{IR-BC}), 
\label{eq:ir-bc}
 \\
  \left. \left(
      1 + \dfrac{1}{2\omega} x\dfrac{d}{dx} 
  \right) B(x) \right|_{x=\Lambda^2} 
  = m_0,
  & & (\mbox{UV-BC}),
\label{eq:uv-bc}
\end{eqnarray}
with $\omega$ being defined by\footnote{
The sign in the definition of $\omega$, Eq.(\ref{eq:def_omega}), is
opposite to the definition in Ref.~\cite{TMSM_ED}.} 
\begin{equation}
  \omega \equiv \frac{1}{2} \left(\frac{D}{2}-1\right).
\label{eq:def_omega}
\end{equation}
The differential equation (\ref{eq:diff1}) is still non-linear and
cannot be solved analytically.
In Ref.~\cite{TMSM_ED}, we discussed similar equation using a
bifurcation method \cite{kn:Atkinson87} in order to deal with the
non-linearity. 
Here in this paper, we use a different method \cite{DSBbook}, in which
Eq.(\ref{eq:diff1}) is replaced by a linearized one
\begin{equation}
  \frac{d}{dx}\left[ x^{D/2} \frac{d}{dx} B \right]
  + \dfrac{2(D-1)(D-2)}{D} \kappa_D \dfrac{x^{D/2-1} B}{x+B_0^2} = 0,
\label{eq:diff2}
\end{equation}
combined with a subsidiary normalization condition
\begin{equation}
  B_0 \equiv B(x=0).
\label{eq:normalization}
\end{equation}
The approximation Eq.(\ref{eq:diff2}) can be shown to work
reasonably well in both high- and low-energy regions.
It has also been used widely in the analysis of the dynamical chiral
symmetry breaking.

It is now easy to solve the differential equation (\ref{eq:diff2}).
Combining it with the IR-BC (\ref{eq:ir-bc}) and the normalization
condition (\ref{eq:normalization}), we find that the solution is given 
in terms of the hypergeometric function,
\begin{equation}
  B(x) = B_0 F(\omega(1+\tilde \nu), \omega(1-\tilde \nu), 
               D/2; -x/B_0^2),
  \qquad
  \tilde\nu \equiv \sqrt{1-\kappa_D/\kappa_D^{\rm crit}}
\label{eq:hyper1}
\end{equation}
for $\kappa_D < \kappa_D^{\rm crit}$, and
\begin{equation}
  B(x) = B_0 F(\omega(1+i\nu), \omega(1-i\nu), 
               D/2; -x/B_0^2),
  \qquad
  \nu \equiv \sqrt{\kappa_D/\kappa_D^{\rm crit}-1}
\label{eq:hyper2}
\end{equation}
for $\kappa_D > \kappa_D^{\rm crit}$, where $\kappa_D^{\rm crit}$ is
given by 
\begin{equation}
  \kappa_D^{\rm crit} = \dfrac{D}{32} \dfrac{D-2}{D-1}.
\label{eq:crit1}
\end{equation}
The critical $\kappa_D^{\rm crit}$ separates chiral symmetric and
broken phases as shown below.
The chiral symmetry breaking takes place for 
$\kappa_D > \kappa_D^{\rm crit}$, while the theory remains chiral
symmetric for  
$\kappa_D < \kappa_D^{\rm  crit}$.
We also define the ``critical coupling'' $\hat g_{\rm crit}$ for
later purpose,
\begin{equation}
  \hat g_{\rm crit}^2 \equiv 
    \dfrac{\kappa_D^{\rm crit}}{C_F \Omega_{\rm NDA}}.
\end{equation}

We note that the $\kappa_D^{\rm crit}$ is larger than the
previous calculation in the Landau gauge,
\begin{equation}
  \kappa_D^{\rm crit} = \dfrac{1}{8} \dfrac{D-2}{D-1},
  \qquad
  \mbox{(Landau gauge)},
\label{eq:crit2}
\end{equation}
where the prescription Eq.(\ref{eq:imp1}) was adopted.
The difference between Eq.(\ref{eq:crit1}) and Eq.(\ref{eq:crit2})
becomes significant for larger $D\gg 4$.
Moreover, Eq.(\ref{eq:crit1}) indicates that the critical coupling
is stronger than the NDA estimate $\kappa_D^{\rm crit} \sim {\cal
  O}(1)$ for $D\gg 4$.
This property can be related to the ``bound state'' problem in $D$
dimensional quantum mechanics.
In order to investigate this issue, we will rewrite the gap equation
into a form of equivalent Schr\"{o}dinger-like equation in
Appendix~\ref{schro}.

We first consider the solution in the subcritical region,
Eq.(\ref{eq:hyper1}). 
In the asymptotic energy region ($x \gg B_0^2$), Eq.(\ref{eq:hyper1})
behaves as
\begin{equation}
  B(x) = B_0 \left[ 
    \tilde c_0 \left(\dfrac{x}{B_0^2}\right)^{-\omega(1-\tilde \nu)}
   +\tilde d_0 \left(\dfrac{x}{B_0^2}\right)^{-\omega(1+\tilde \nu)}
   + {\cal O}\left(\left(\dfrac{x}{B_0^2}\right)^{-\omega(1-\tilde \nu)-1}\right)
  \right],
\label{eq:asympt1}
\end{equation}
with $\tilde c_0$ and $\tilde d_0$ being given by
\begin{displaymath}
  \tilde c_0 \equiv
    \dfrac{\Gamma(D/2) \Gamma(2\omega\tilde\nu)}
          {\Gamma(\omega(1+\tilde\nu))\Gamma(1+\omega(1+\tilde\nu))},
  \quad
  \tilde d_0 \equiv
    \dfrac{\Gamma(D/2) \Gamma(-2\omega\tilde\nu)}
          {\Gamma(\omega(1-\tilde\nu))\Gamma(1+\omega(1-\tilde\nu))}.
\end{displaymath}
Plugging Eq.(\ref{eq:asympt1}) into the UV-BC Eq.(\ref{eq:uv-bc}), 
we obtain
\begin{equation}
  \frac{1}{2}(1+\tilde \nu) \tilde c_0
  B_0\left(\dfrac{\Lambda^2}{B_0^2}\right)^{-\omega(1-\tilde\nu)}
  \simeq m_0.
\label{eq:uv-bc2}
\end{equation}
The non-trivial solution $B_0 \ne 0$
exists only when $m_0 \ne 0$, i.e., 
the dynamical chiral symmetry breaking does not occur in the
subcritical region $\kappa_D < \kappa_D^{\rm crit}$.

The situation differs significantly for 
$\kappa_D > \kappa_D^{\rm crit}$. 
The high energy behavior of Eq.(\ref{eq:hyper2}) is 
\begin{equation}
  B(x) = B_0 \left[ 
    c_0 \left(\dfrac{x}{B_0^2}\right)^{-\omega(1-i\nu)}
   +d_0 \left(\dfrac{x}{B_0^2}\right)^{-\omega(1+i \nu)}
   + {\cal O}\left( \left(\dfrac{x}{B_0^2}\right)^{-\omega-1}\right)
  \right],
\label{eq:asympt2}
\end{equation}
with 
\begin{displaymath}
  c_0 \equiv
    \dfrac{\Gamma(D/2) \Gamma(2i\omega\nu)}
          {\Gamma(\omega(1+i\nu))\Gamma(1+\omega(1+i\nu))},
  \qquad
  d_0 \equiv c_0^*.
\end{displaymath}
Unlike the solution at $\kappa_D < \kappa_D^{\rm crit}$,
Eq.(\ref{eq:asympt2}) is an oscillating function.
Actually, Eq.(\ref{eq:asympt2}) can be written as
\begin{equation}
  B(x) \simeq 2|c_0| B_0 \left(\dfrac{x}{B_0^2}\right)^{-\omega}
         \sin\left[\theta + \omega\nu \ln \frac{x}{B_0^2}\right],
\label{eq:sol_bx}
\end{equation}
with $\theta$ being given by $e^{2i\theta} = -c_0/d_0$.
The UV-BC Eq.(\ref{eq:uv-bc}) thus reads
\begin{equation}
\sqrt{1+\nu^2} |c_0| B_0 \left(\dfrac{\Lambda^2}{B_0^2}\right)^{-\omega}
         \sin\left[\theta + \omega\nu \ln \frac{\Lambda^2}{B_0^2}
                          +\tan^{-1}\nu 
             \right] = m_0.
\label{eq:crit-uv-bc}
\end{equation}
Non-trivial solutions $B_0 \ne 0$ exist even in the chiral limit
$m_0=0$ of Eq.(\ref{eq:crit-uv-bc}):
\begin{equation}
  B_0 \simeq  C \Lambda \exp\left[
    -\dfrac{n\pi}{(D/2-1)\nu} 
  \right],
\label{eq:sol_b0}
\end{equation}
with $n$ being a positive integer describing the number of nodes in $B(x)$. 
The $n=1$ solution corresponds to the nodeless one. 
Here the factor $C$ is given by
$C \equiv \exp((\theta + \tan^{-1}\nu)/((D/2-1)\nu))$,
which remains finite in the $\nu\rightarrow 0$ limit 
($\nu\equiv \sqrt{\kappa_D/\kappa_D^{\rm crit}-1}$).

\end{widetext} 

Although we found infinite number of solutions in
Eq.(\ref{eq:sol_b0}) labeled by $n$, these solutions may correspond 
unstable vacua.
We need to evaluate the vacuum energy in order to find the true vacuum
with minimum energy.
We thus compare energies of different vacua ($n=1,2,\cdots$)
by using the Cornwall-Jackiw-Tomboulis (CJT)
effective potential \cite{kn:CJT}.
Within the improved ladder approximation,
the CJT potential is given by
\begin{widetext} 
\begin{eqnarray}
\lefteqn{
  - \dfrac{V_{\rm CJT}[B,m_0,\Lambda)}
          {\eta N N_f \Omega_{\rm NDA}}
  =
   \int_0^{\Lambda^2} dx x^{D/2-1} \left\{
       \frac{1}{2} \ln\left(1+\dfrac{B^2(x)}{x}\right)
      -\dfrac{B^2(x)-m_0B(x)}{x + B^2(x)}
      \right\}
}  \hspace*{2.5cm} \nonumber\\
  & & \quad
      + \frac{1}{2} C_F 
        \int_0^{\Lambda^2} dx x^{D/2-1} \! \dfrac{B(x)}{x+B^2(x)}
        \int_0^{\Lambda^2} dy y^{D/2-1} \! \dfrac{B(y)}{x+B^2(y)}
        K_B(x,y).
\label{eq:cjt}
\end{eqnarray}
It is easy to show that the stationary condition of the CJT potential
$\delta V_{\rm CJT}/\delta B = 0$ is identical to the gap equation
Eq.(\ref{eq:gapeq1}). 

Using a scaling technique described in Ref.~\cite{GM87} we can evaluate the
CJT effective potential,
\begin{equation}
  - \dfrac{D}{\eta N N_f \Omega_{\rm NDA}}
    V_{\rm CJT}[B_{\rm sol},m_0,\Lambda)
  = \Lambda^D \ln\left(1+\dfrac{B_\Lambda^2}{\Lambda^2}\right)
   +\dfrac{D}{4(D-1)\kappa_D} \Lambda^{D-2}(B_\Lambda-m_0)m_0,
\label{eq:vac-energy}
\end{equation}
\end{widetext} 
where
\begin{equation}
 B_\Lambda \equiv B_{\rm sol}(x=\Lambda^2),
\end{equation}
with $B_{\rm sol}$ being the solution of the gap equation 
Eq.(\ref{eq:gap2}).
We note that the vacuum energy $V_{\rm CJT}$ is a decreasing function of
$|B_\Lambda|$ in the chiral limit $m_0=0$.
Combining Eq.(\ref{eq:sol_b0}) and Eq.(\ref{eq:sol_bx}),
it is easy to obtain
\begin{equation}
  |B_\Lambda| \simeq \frac{2 |c_0| \nu}{\sqrt{1+\nu^2}}
  \Lambda \left(\dfrac{B_0}{\Lambda}\right)^{D/2}. 
\end{equation}
We thus find that the vacuum energy is a decreasing function of $B_0$.
It is now straightforward to show that the vacuum with minimum
energy
in the chiral limit $m_0=0$ corresponds to the $n=1$
solution (largest $B_0$)\footnote{
It should be mentioned that the $n=1$ solution is not an absolute
minimum, but a saddle point of the CJT potential.
However, this fact does not indicate the instability of this vacuum, since
negative curvature in the CJT potential does not necessarily imply the
existence of a tachyonic mode. 
We will discuss another potential with possibly better perspective in
Appendix~\ref{app:eff_pot}. 
}
in Eq.(\ref{eq:sol_b0}) at the
super-critical $\kappa_D > \kappa_D^{\rm crit}$.

We thus obtain the scaling relation for 
$\kappa_D \simeq \kappa_D^{\rm crit}$, 
\begin{equation}
  B_0 \propto \Lambda \exp\left[
    \dfrac{-\pi}{(D/2-1)\sqrt{\kappa_D/\kappa_D^{\rm crit}-1}}
  \right].
\label{eq:sol-sd1}
\end{equation}
As pointed out in Ref.~\cite{TMSM_ED}, 
the chiral phase transition Eq.(\ref{eq:sol-sd1}) is an
essential-singularity type (as a result of ``conformal phase
transition'' \cite{MY97}), which enables us to obtain a hierarchy  
between the cutoff $\Lambda$ and the dynamical mass $B_0$ in a model
with $\kappa_D$ sufficiently close to the critical $\kappa_D^{\rm  crit}$. 
Actually, it is easy to realize a hierarchy of ${\cal O}(10)$ level 
without any fine-tuning, e.g., 
$\Lambda/B_0\simeq 12$ in the $SU(3)$ gauge theory with  $N_f=5$ and
$D=4+2$.

Note, however, that $\kappa_D$ is not an adjustable free parameter, but
a definite number once the model is set up.
We are thus not able to obtain arbitrarily large hierarchy
$\Lambda/B_0$ in models standing at the UV-FP\@.
In order to clarify the point, it is useful to translate
$\kappa_D^{\rm crit}$ to critical number of flavors $N_f^{\rm crit}$
by noting that $\kappa_D$ depends on $N_f$,
\begin{equation}
  N_f^{\rm crit}
  = \dfrac{3}{\eta T_R} \left[ 
      \dfrac{26-D}{6} C_G 
      - \dfrac{C_F}{\frac{2}{D-4}+1}\dfrac{32(D-1)}{D(D-2)}
    \right],
\label{eq:critNf}
\end{equation}
where we have used Eq.(\ref{eq:betafun}) and Eq.(\ref{eq:uvfp1}).
($\kappa_D < \kappa_D^{\rm crit}$ corresponds to $N_f < N_f^{\rm crit}$.)
For $SU(3)$ gauge theories, the critical $N_f$
is evaluated as, 
\begin{equation}
  N_f^{\rm crit} \simeq 
  \begin{cases}
    4.2,  & \text{for $D=4+2$}, \\
    1.8,  & \text{for $D=4+4$}, \\
    0.8,  & \text{for $D=4+6$}. \\
  \end{cases}
\label{eq:Ncri_MS}
\end{equation}
The largest hierarchy in $SU(3)$ gauge theories
is then obtained when $N_f=5$, $N_f=2$ and $N_f=1$ 
in $D=4+2$, $D=4+4$ and $D=4+6$ dimensions, respectively.

In order to estimate the uncertainties coming from the
compactification scale $R^{-1}$, we next consider the effects of an
infrared (IR) cutoff $M_0^2 \sim R^{-2}$.
It is shown in Appendix~\ref{ir-cutoff} that the critical $\kappa_D$
is a function of $M_0^2$ under the presence of IR cutoff $M_0^2$,
\begin{equation}
  \kappa_D^{\rm crit}(M_0^2) 
    = \kappa_D^{\rm crit}(M_0^2=0) \left( 1 + \nu_c^2(M_0^2) \right),
\end{equation}
with $\kappa_D^{\rm crit}(M_0^2=0)$ being the critical point without IR
cutoff and given by Eq.(\ref{eq:crit1}).
$\nu_c(M_0^2)$ is a solution of
\begin{equation}
  2\tan^{-1}\nu_c + \omega \nu_c \ln\dfrac{\Lambda^2}{M_0^2} = \pi. 
\end{equation}
For $\Lambda \gg  M_0$, $\nu_c(M_0^2)$ is thus given by
\begin{equation}
  \nu_c(M_0^2) \simeq \dfrac{\pi}{2+\omega\ln\dfrac{\Lambda^2}{M_0^2}}
               \ll 1.
\end{equation}
The difference between $\kappa_D^{\rm crit}(M_0^2 \ne 0)$ and 
$\kappa_D^{\rm crit}(M_0^2 = 0)$ can be neglected for sufficiently
large UV cutoff $\Lambda$.

On the other hand, the scaling behavior of $B(M_0^2)$ coincides with
Eq.(\ref{eq:sol-sd1}) (see also Eq.(\ref{eq:sol_b0})) for $M_0 \ll
B(M_0^2) \ll \Lambda$, while 
\begin{equation}
  B(M_0^2) \propto M_0 \dfrac{\sqrt{\nu - \nu_c(M_0^2)}}{\nu_c(M_0^2)},
\label{eq:scalingM_0}
\end{equation}
for $B(M_0^2) \ll M_0$.
Eq.(\ref{eq:scalingM_0}) indicates mean-field type scaling.
We thus need a fine-tuning of model in order to realize hierarchy 
between the fermion mass $B(M_0^2)$ and the IR cutoff 
$M_0$($\sim R^{-1}$).

If the effects of the compactification scale $R^{-1}$ can be mimicked
by the IR cutoff $M_0$, 
the result Eq.(\ref{eq:scalingM_0}) implies that the dynamical fermion
mass $B_0$ cannot be made extremely smaller than $R^{-1}$  without
severe fine-tuning.
It is, however, relatively easy to achieve ${\cal O}(10)$ level
hierarchy between UV cutoff $\Lambda$ and $B_0$ as we have discussed
before.

\begin{figure*}[tbp]
  \begin{center}
    \resizebox{0.45\textwidth}{!}{
      \includegraphics{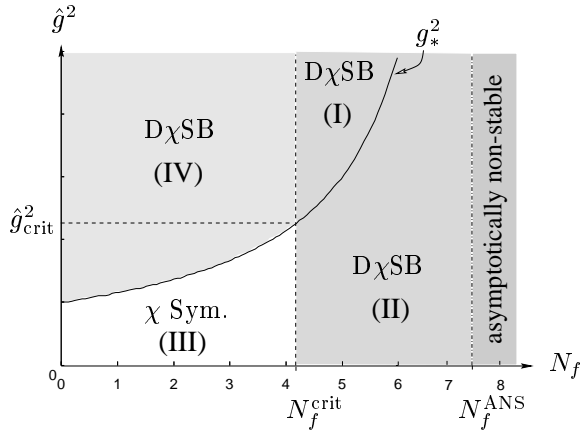}}
    \caption{
      The phase diagram of $SU(3)$ gauge theory in
      $D=4+2$ dimensions.
      While the theory remains 
      chiral symmetric for $N_f < N_f^{\rm crit}$ with $\hat g < g_*$, 
      the dynamical chiral
      symmetry breaking takes place in the entire region of $\hat g$
      with a sufficiently large cutoff $\Lambda$ for
      $N_f^{\rm crit} < N_f < N_f^{\rm ANS}=7.5$.
      The theory becomes asymptotically non-stable without UV-FP for $N_f
      > N_f^{\rm ANS}$.
}
    \label{fig:phase_diagram}
  \end{center}
\end{figure*}

\section{Phase structure}

We have so far discussed the chiral phase transition assuming that
the gauge coupling $\hat g$ is standing at the UV-FP\@.
We found the critical $N_f^{\rm crit}$, Eq.(\ref{eq:critNf}).
The chiral symmetry breaking takes place only when 
$N_f>N_f^{\rm crit}$ at the UV-FP\@.

In particle models with extra dimensions, however, the gauge coupling
$\hat g$ is not necessarily on its UV-FP, $g_*$.
The bulk QCD coupling of the ACDH scenario of the top condensate, for
example, is shown to be below its UV-FP\@. 
In this section, we therefore try to draw more concrete picture of the
phase diagram including the gauge coupling strength apart from the
UV-FP\@. 
It is illuminating to discuss the phase diagram in the 
$\hat g^2$-$N_f$ plane.

In the following analysis, particular interests are paid for $SU(3)$
gauge theories in $D=4+2$ dimensions, in which the critical $N_f$ is
evaluated as $N_f^{\rm crit}\simeq 4.2$. 
Since the chiral symmetry breaking is our main concern, we take the
chiral limit $m_0=0$ in this section.

Before starting the detailed numerical analysis, we first discuss
the qualitative picture of the phase diagram
(Fig.\ref{fig:phase_diagram}) 
by using the result of the gap equation at the UV-FP\@.

\begin{figure}[htbp]
  \begin{center}
    \resizebox{0.45\textwidth}{!}{
      \includegraphics{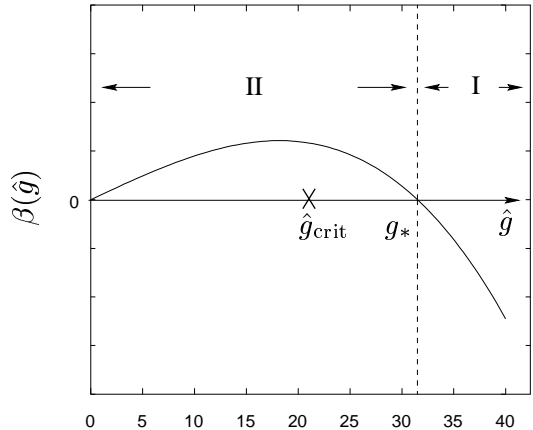}}
    \caption{The beta function of $SU(3)$ gauge theory with $N_f=6$ in
      $D=4+2$ dimensions. 
      The UV-FP is above the critical coupling $\hat g_{\rm crit}$. 
}
    \label{fig:beta_nf6}
  \end{center}
\end{figure}

Let us start with the case $N_f^{\rm ANS}>N_f>N_f^{\rm crit}$.
The typical behavior of the beta function of the dimensionless gauge
coupling $\hat g$ is depicted in Fig.~\ref{fig:beta_nf6} with this $N_f$.
The UV-FP is above the critical coupling $\hat g_{\rm crit}$.
The bulk fermion then acquires its dynamical mass
proportional to the cutoff $\Lambda$, Eq.(\ref{eq:sol-sd1}), even at
the UV-FP\@.
We therefore expect the chiral symmetry breaking takes place in the
strongly coupled regime $\hat g > g_*$ (region I in
Fig.~\ref{fig:phase_diagram}). 
Since the gauge coupling $\hat g$ approaches quickly to its UV-FP
in the asymptotic region, 
the coupling
exceeds its critical value for sufficiently large energy scale even in
the weakly coupled region (region II).
It is then expected that dynamical chiral symmetry breaking occurs
even in region II for sufficiently large cutoff.
In order to keep the fermion mass finite, the cutoff $\Lambda$ needs
to be finite.

\begin{figure}[htbp]
  \begin{center}
    \resizebox{0.45\textwidth}{!}{
      \includegraphics{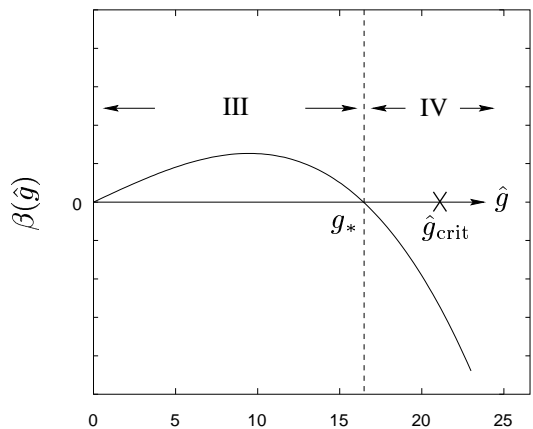}}
    \caption{The beta function of $SU(3)$ gauge theory with $N_f=2$ in
      $D=4+2$ dimensions. 
      The UV-FP is below the critical coupling $\hat g_{\rm crit}$. 
}
    \label{fig:beta_nf2}
  \end{center}
\end{figure}

The situation should be substantially different for 
$N_f < N_f^{\rm crit}$. (See Fig.~\ref{fig:beta_nf2}.) 
In this case, the fermion remains massless at the UV-FP
$\hat g = g_*$ or $\Q =0$ with $\Q$ being the scale of the gauge theory
defined in Eq.(\ref{eq:msbar6}).
The gauge coupling $\hat g$, therefore, does not exceeds its critical
value $\hat g_{\rm crit}$ in the weakly coupled regime $(\Q)^2<0$ 
(region III). 
On the other hand, the gauge coupling in the strongly coupled phase
$(\Q)^2>0$ (region IV) becomes extremely strong at the infrared region
$\mu^2 \simeq (\Q)^2$ and the coupling becomes stronger than its critical
value $\hat g_{\rm crit}$.  
It is therefore expected that the fermion acquires its dynamical mass
$M^2 \sim (\Q)^2$ in region IV, while the theory remains chiral symmetric 
in region III\@.
The cutoff $\Lambda$ can be arbitrarily large in the analysis of the
gap equation for $N_f < N_f^{\rm crit}$. 

If we neglect effects of the weak gauge interactions, the minimal ACDH
scenario corresponds to the $D=4+2$ dimensional $SU(3)$ (QCD) gauge
theory with $N_f=2$ and $\hat g < g_*$.
Eq.(\ref{eq:Ncri_MS}) implies $N_f^{\rm crit}\simeq 4.2$ in this case
and Fig.~\ref{fig:phase_diagram} shows the model is in its chiral
symmetric phase (region III).
The QCD gauge coupling is thus not strong enough to trigger the chiral
condensate in this scenario no matter how large the cutoff $\Lambda$
is. 
The model thus needs to be modified in order to explain the
electroweak symmetry breaking.
For example, the $D=4+4$ dimensional version of the ACDH scenario is
shown to be in the chiral symmetry breaking phase (region II)
($N_f^{\rm crit}\simeq 1.8$) and it
may explain the mass of weak gauge bosons. 

It is also worth pointing out that the region IV in the phase diagram
(Fig.~\ref{fig:phase_diagram}) may open an interesting possibility in
the model buildings of the electroweak symmetry breaking.
The region IV is very interesting because we can formally take
$\Lambda\rightarrow \infty$ limit in the analysis of the gap equation.
The low energy predictions of the models in this phase are thus
insensitive to the physics around the UV cutoff.
One of the possibilities is an idea of  ``bulk technicolor'' model.
The phenomenology of this scenario will be discussed in a separated
publication. 

\begin{figure}[htbp]
  \begin{center}
    \resizebox{0.45\textwidth}{!}{
      \includegraphics{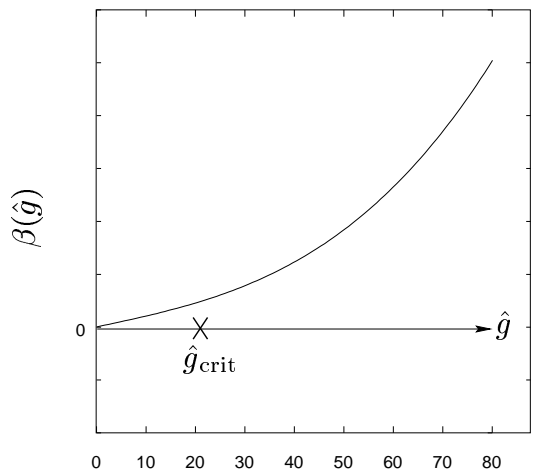}}
    \caption{The beta function of $SU(3)$ gauge theory with $N_f=8$ in
      $D=4+2$ dimensions. 
      The theory becomes asymptotically non-stable (ANS).
}
    \label{fig:beta_nf8}
  \end{center}
\end{figure}

Finally, we consider the case with $N_f>N_f^{\rm ANS}$, where 
the theory becomes asymptotically non-stable. (See
Fig.~\ref{fig:beta_nf8}).
The gauge coupling $\hat g$ grows without a bound and exceeds 
$\hat g_{\rm crit}$ in the high energy region.
The chiral symmetry is then expected to be broken spontaneously for
sufficiently large cutoff $\Lambda$.

In order to confirm these expectations, however,
we need to investigate the gap equation with the gauge coupling
strength away from its fixed point, which we will perform in the
following subsections.

\subsection{$N_f < N_f^{\rm crit}$ \label{sec:weakly_int}}

Let us first consider the chiral symmetry breaking in the strongly
interacting phase $(\Q)^2 >0$ with $N_f < N_f^{\rm crit}$ (region IV).

Note here that Eq.(\ref{eq:msbar6}) has a difficulty associated
with the (tachyonic) pole singularity for $(\Q)^2>0$ (strongly coupled
phase). 
Such a singularity appears only when the gauge coupling becomes
extremely strong.  So it should be an artifact of one-loop
approximation of the beta function.
In order to avoid the difficulty, 
we make an ansatz of the dimensionless gauge coupling $\hat g^2 \equiv
\mu^2 g_D^2$ in the infrared region $\mu^2<\tau (\Q)^2$ with $\tau > 1$, 
\begin{equation}
  \hat g^2(\mu^2) \Omega_{\rm NDA}
 = \dfrac{1}{-2b'} \dfrac{1}{(\tau-1)^2} \left[
     \tau^2 -\dfrac{\mu^2}{(\Q)^2}
   \right], 
\label{eq:IRregulator}
\end{equation}
for $\mu^2 < \tau (\Q)^2$.
The form Eq.(\ref{eq:IRregulator}) is a linear function of $\mu^2$ and 
is taken so as to make
$\hat g^2(\mu^2)$ and its first derivative continuous at $\mu^2=\tau (\Q)^2$.
The regulator $\tau (\Q)^2$ is chosen to make the coupling at
$g_D^2(\mu^2=0)$ sufficiently large:
\begin{eqnarray}
  \hat g_{\rm IR}^2 \Omega_{\rm NDA}
  &=& \left. \hat g^2(\mu^2) \Omega_{\rm NDA} \right|_{\mu^2 = 0}
      \nonumber \\ 
  &=& \dfrac{1}{-2b'} \dfrac{\tau^2}{(\tau-1)^2}
                                      \sim {\cal O}(1).
\end{eqnarray}

\begin{figure}[htbp]
  \begin{center}
    \resizebox{0.45\textwidth}{!}{
     \includegraphics{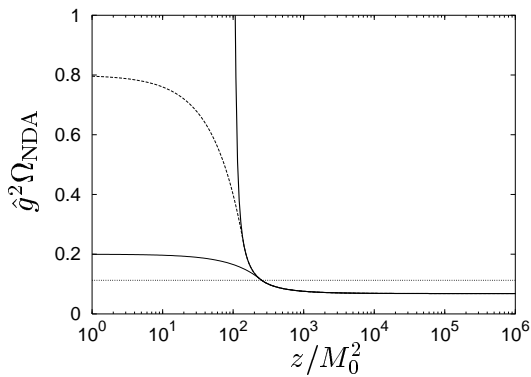}}
    \caption{The momentum dependence of the running gauge coupling
     in the strongly coupled phase of the $D=4+2$ $SU(3)$ gauge theory
     ($N_f=2$, $(\Q)^2=10^2 M_0^2$). 
     The infrared regularization $g_{\rm IR}$ is taken to be $0.2$ (solid line)
     and $0.8$ (dashed line). 
     The dotted line is the critical coupling 
    ($\hat g_{\rm crit}^2 \Omega_{\rm NDA}=9/80$).
    }
    \label{fig:gd2msbar}
  \end{center}
\end{figure}

\begin{figure}[htbp]
  \begin{center}
    \resizebox{0.45\textwidth}{!}{
    \includegraphics{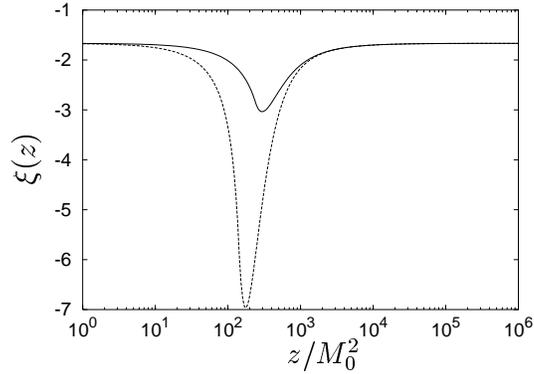}}
    \caption{The momentum dependence of the nonlocal gauge fixing
    parameter in the strongly coupled phase of the $D=4+2$ $SU(3)$ gauge
    theory ($N_f=2$, $(\Q)^2=10^2 M_0^2$).
    The infrared regularization $g_{\rm IR}$ is taken to be $0.2$ (solid line) 
    and $0.8$ (dashed line).}
    \label{fig:xi_msbar}
  \end{center}
\end{figure}

The behaviors of the gauge coupling of $D=4+2$ $SU(3)$ $N_f=2$ gauge theory
are shown in Fig.~\ref{fig:gd2msbar} for positive $(\Q)^2=10^2 M_0^2>0$.
Each lines correspond to the non-reguralized, and the reguralized $\hat g$
with $\hat g_{\rm IR} \Omega_{\rm NDA}=0.2$ and $0.8$. 
The non-reguralized $\hat g$ diverges at the scale $(\Q)^2$.
Corresponding behaviors of the non-local gauge fixing parameter $\xi$
are shown in Fig.\ref{fig:xi_msbar}.

\begin{figure}[htbp]
  \begin{center}
    \resizebox{0.45\textwidth}{!}{
    \includegraphics{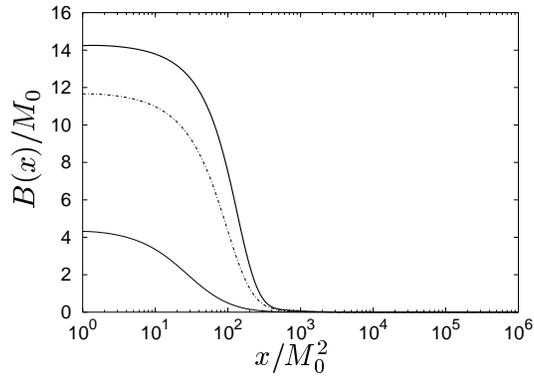}}
    \caption{The solution of the gap equation of $D=4+2$ dimensional
    $SU(3)$ gauge theory with $N_f=2$. $(\Q)^2=10^2 M_0^2$ and
    $\Lambda^2 = 10^6 M_0^2$ are assumed.
    The results with various choice of the infrared regulator 
    $\hat g_{\rm IR}^2\Omega_{\rm NDA}=0.2$ (solid line), 
    0.5 (dashed line), 0.8 (bold line) are depicted.
}
    \label{fig:bb_msbar}
  \end{center}
\end{figure}

The gap equation is solved in a numerical manner by adopting
the recursion method \cite{TMSM_ED}.
We also introduce an infrared cutoff $M_0$.
In the following numerical analysis,  $(\Q)^2=10^2 M_0^2$ and
$\Lambda^2=10^6 M_0^2$ are assumed. 

The numerical solution of the gap equation $B(x)/M_0$ is shown in
Fig.\ref{fig:bb_msbar} for various choice of the infrared regulator
$\hat g_{\rm IR}^2 \Omega_{\rm NDA}$.
We find 
\begin{equation}
  B(M_0^2) \sim \Q = \Lambda \sqrt{1-\dfrac{g_*^2}{\hat g^2(\mu=\Lambda)}}
\end{equation}
for sufficiently large $\hat g_{\rm IR}$ as we have expected before.
The solution is insensitive to the choice of the UV cutoff
$\Lambda^2$.
We can thus formally define a continuum limit (infinite cutoff
limit), which implies that the low-energy physics becomes {\it
insensitive to the details of the physics around cutoff}. 
It should be emphasized that this phase may be useful within the
framework of local field theory, no matter whatever physics may exist
behind the UV cutoff.

\begin{figure}[htbp]
  \begin{center}
    \resizebox{0.5\textwidth}{!}{
    \includegraphics{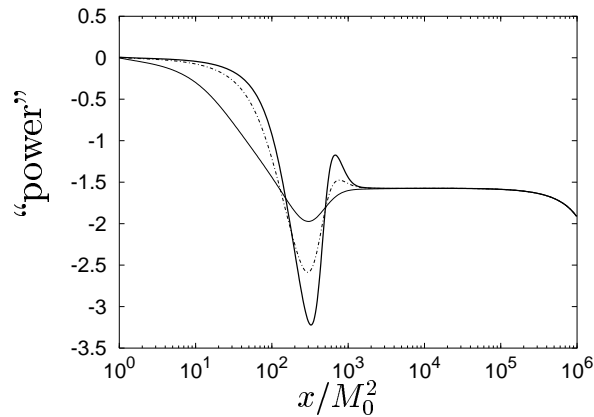}}
    \caption{The ``power'' behavior of the numerical solution.
    The infrared regulator is taken as $\hat g_{\rm IR}^2\Omega_{\rm
    NDA}=0.2$ (solid line), 0.5 (dashed line), 0.8 (bold line).} 
    \label{fig:pw_msbar}
  \end{center}
\end{figure}

We next consider the high-energy ($x \gg (\Q)^2 \sim B^2$) behavior of
the mass function.
In the asymptotic energy region, the gauge coupling strength
approaches very quickly to its UV-FP\@.
We therefore expect that the asymptotic solution satisfies the
differential equation Eq.(\ref{eq:diff1}) and the UV-BC
Eq.(\ref{eq:uv-bc}) which were derived originally at the UV-FP\@.
On the other hand, the infrared behavior of the solution should
be substantially different from the solution at the UV-FP\@.
We therefore do not adopt the IR-BC Eq.(\ref{eq:ir-bc}).

Eq.(\ref{eq:diff1}) can be approximated further for $x\gg B^2$,
\begin{eqnarray}
\lefteqn{  \frac{d}{dx}\left[ x^{D/2} \frac{d}{dx} B \right]
} \hspace*{2mm} \nonumber \\ &&
  + \dfrac{2(D-1)(D-2)}{D} \kappa_D x^{D/2-2} B = 0.
\label{eq:diff3}
\end{eqnarray}
It is easy to solve Eq.(\ref{eq:diff3}) and Eq.(\ref{eq:uv-bc}).
We find the asymptotic solution is given by
\begin{equation}
  B(x) \propto
   \left(\dfrac{x}{\Lambda^2}\right)^{-\omega(1+\tilde\nu)}\!\!\!\!
  -\dfrac{1-\tilde\nu}{1+\tilde\nu}
   \left(\dfrac{x}{\Lambda^2}\right)^{-\omega(1-\tilde\nu)},
\label{eq:asympt_sol}
\end{equation}
with 
$\tilde\nu \equiv \sqrt{1-\kappa_D/\kappa_D^{\rm crit}}
           =  \sqrt{1-g_*^2/\hat g_{\rm crit}^2}
$.
The second term is negligible for $\Lambda^2 \gg x$.
We thus expect that the mass function $B$ behaves as
\begin{equation}
  B(x) \propto
   \left(\dfrac{x}{\Lambda^2}\right)^{-\omega(1+\tilde\nu)} \label{asympt3}
\end{equation}
in the energy region $\Lambda^2 \gg x \gg (\Q)^2$ if a non-trivial 
$B\ne 0$ solution exists.

In order to confirm the above expectation,
we next plot
the ``power'' behavior of the numerical solution of the mass function 
\begin{equation}
  \mbox{``power''} = \dfrac{x}{B(x)}\dfrac{dB(x)}{dx}
\end{equation}
in Fig.\ref{fig:pw_msbar}.
The asymptotic behavior\footnote{
The behavior  near the cutoff $x \simeq \Lambda^2$ in Fig.\ref{fig:pw_msbar}
is an artifact due to the sharp cutoff introduced in the analysis of the 
gap equation~\cite{kn:NT89}.
}
Eq.(\ref{asympt3}) 
is consistent with $\mbox{``power''} \simeq -1.6$ in Fig.\ref{fig:pw_msbar},
which agrees well with the expected value $-\omega(1+\tilde\nu)$
where $\omega=1, \tilde\nu \simeq 0.6$ 
for the $D=4+2$ dimensional $SU(3)$ gauge theory with $N_f=2$.

We also note that the ``power'' is related to the anomalous dimension
of the fermion mass $\gamma_m$ as 
$\mbox{``power''}=\gamma_m/2-(D/2-1)$~\cite{TMSM_ED}. 
The anomalous dimension $\gamma_m$ in the asymptotic region is then
given by
\begin{equation}
  \gamma_m = \left(\frac{D}{2}-1\right)
             \left( 1-\sqrt{ 1 - g_*^2 /  \hat g_{\rm crit}^2 }\right).
\end{equation}

\begin{figure}[htbp]
  \begin{center}
    \resizebox{0.45\textwidth}{!}{
    \includegraphics{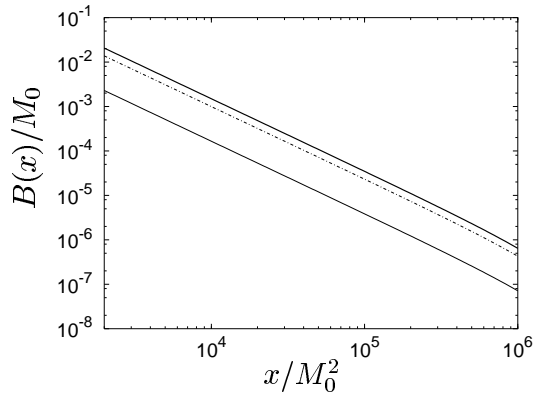}}
    \caption{The asymptotic behavior of the mass function for $\hat 
      g_{IR}^2\Omega_{\rm NDA}=0.2$ (solid line), 0.5 (dashed line), 
      0.8 (bold line).}
    \label{fig:bb_msbar2}
  \end{center}
\end{figure}

We next examine the absolute magnitude of the mass function in the
asymptotic region.
For such a purpose, we show a log-log plot of the mass function
in the asymptotic region. (See Fig.\ref{fig:bb_msbar2}. )
We find that the asymptotic mass function becomes insensitive to the
infrared regulator $\hat g_{\rm IR}$ if we take $\hat g_{\rm IR}^2
\Omega_{\rm NDA} \gtrsim 0.5$.

The infrared behavior of the solution depends significantly on the
choice of the infrared regulator $\hat g_{\rm IR}^2$. 
The infrared behavior is therefore not enough trustworthy in this
calculation. 
It should be emphasized, however, that the ultraviolet behavior is
relatively insensitive to the choice of infrared regulator.

We have so far discussed the case where $\Q$ is sufficiently large
compared with the IR cutoff $M_0$ and found the dynamical chiral
symmetry breaking $B(M_0^2)\sim \Q$.
The situation differs substantially for $\Q \ll M_0$, where
the gauge coupling $\hat g$ cannot exceeds its critical value of
the chiral phase transition $\hat g_{\rm crit}$.
Actually, the dynamical chiral symmetry breaking does not take place
for $\Q \ll M_0$.

Similar analysis is also performed in the weakly interacting phase
 $(\Q)^2<0$ (region III).
As we expected before, 
we find no signal of chiral symmetry breaking in this phase with $N_f
 < N_f^{\rm crit}$. 

\subsection{$N_f > N_f^{\rm crit}$}

We next discuss gauge theories with $N_f > N_f^{\rm crit}$ (regions I
and II).
In these models, the UV-FP of the gauge coupling is strong
enough to trigger the dynamical chiral symmetry breaking.
We therefore expect that the bulk fermion acquires its dynamical mass
even in the weakly coupled phase $(\Q)^2<0$ (region II).
In order to confirm this expectation, we next investigate the gap
equation with negative $(\Q)^2$, i.e., weakly coupled phase with 
$N_f > N_f^{\rm crit}$. 

\begin{figure}[htbp]
  \begin{center}
    \resizebox{0.45\textwidth}{!}{
    \includegraphics{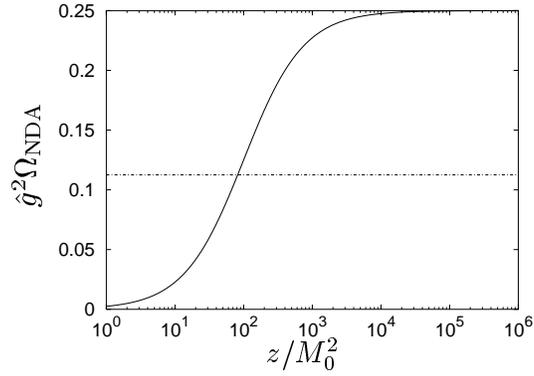}}
    \caption{The running of gauge coupling strength. $D=4+2$, $N=3$,
      $N_f=6$, $(\Q)^2/M_0^2=-100$.
     The dashed line is the critical coupling 
    ($\hat g_{\rm crit}^2 \Omega_{\rm NDA}=9/80$).}
    \label{fig:gd2n_msbar}
  \end{center}
\end{figure}

Fig.\ref{fig:gd2n_msbar} shows the behavior of the $\overline{\rm MS}$ 
gauge coupling of $SU(3)$ gauge theory with $D=4+2$, $N_f=6$.
The scale $(\Q)^2$ is taken as $(\Q)^2/M_0^2=-100$ in this figure.
The gauge coupling approaches its UV-FP 
$g_*^2 \Omega_{\rm NDA}=0.25$ very quickly.
We expect that the dynamical chiral symmetry breaking occurs when the
gauge coupling strength exceeds its critical value:
\begin{equation}
  C_F \hat g^2 \Omega_{\rm NDA} > \kappa_D^{\rm crit} = \frac{3}{20},
\label{eq:dxsbcond1}
\end{equation}
which is actually satisfied for $\mu^2 \gtrsim |(\Q)^2|$.

\begin{figure}[htbp]
  \begin{center}
    \resizebox{0.45\textwidth}{!}{
    \includegraphics{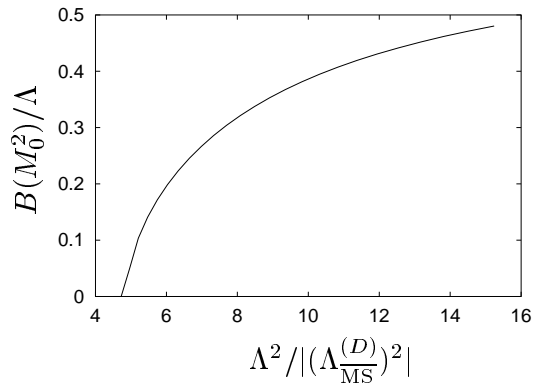}}
    \caption{The scaling behavior of $B(M_0^2)/\Lambda$ as a function
    of $\Lambda^2/|(\Q)^2|$ in the $D=4+2$ dimensional $SU(3)$ theory
    with $N_f=6$. 
    $(\Q)^2/M_0^2=-100$ is assumed.}
    \label{fig:sd6-msbar1n-6.eps}
  \end{center}
\end{figure}

It is straightforward to solve the gap equation numerically.
We find that the dynamical chiral symmetry breaking actually occurs
when the cutoff $\Lambda$ is large enough, 
$\Lambda^2 \gtrsim 5|(\Q)^2|$,  with above mentioned parameters.
Fig.\ref{fig:sd6-msbar1n-6.eps} shows the scaling behavior of
$B(M_0^2)/\Lambda$ as a function of $\Lambda^2/|(\Q)^2|$.
It should be emphasized that the cutoff $\Lambda$ can be determined in 
this case once $B(M_0^2)$ and $(\Q)^2$ are fixed.

\begin{figure}[htbp]
  \begin{center}
    \resizebox{0.45\textwidth}{!}{
    \includegraphics{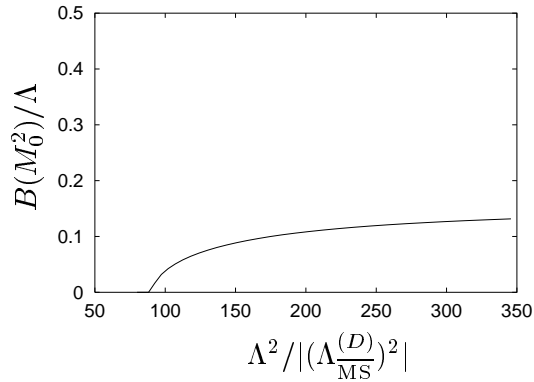}}
    \caption{The scaling behavior with $N_f=5$.
      Other parameters are the same as
      Fig.~\ref{fig:sd6-msbar1n-6.eps}.
}
    \label{fig:sd6-msbar1n-5.eps}
  \end{center}
\end{figure}

The top-mode standard model (TMSM) with extra dimensions would be one of the
most important applications of this phase.
As we have discussed before, we note that the cutoff $\Lambda^2$ is, in
principle, a calculable parameter in the analysis of the gap equation.
Once the cutoff $\Lambda$ is 
determined, 
we can evaluate the decay constant of the NG boson (VEV of Higgs) $v$
by using the Pagels-Stokar formula \cite{kn:PS79}.\footnote{
It is also possible to evaluate $v$ by adopting the
Bardeen-Hill-Lindner (BHL) type compositeness condition \cite{BHL} in
the renormalization group analysis, however without freedom to adjust $\Lambda$.
The decay constant $v$ is given by $v=\sqrt{2} m_t / y_t$ with $y_t$
being the Yukawa coupling satisfying the BHL condition.}
We are thus able to test the scenario by comparing the calculated $v$
with the actual value $v\simeq 250$GeV.
In other words, we can ``predict'' the UV cutoff once we fix the VEV
$v$ to the actual value.
This property is due to the fact that the top-condensate is driven
solely by the bulk QCD gauge coupling which cannot be adjusted
arbitrarily in this scenario. 
It is therefore completely different from the original version of the 
top-mode standard model, where the four-fermion coupling is introduced
as an adjustable free parameter.

This fact is in sharp contrast to the renormalization group analysis
of ACDH, where the cutoff $\Lambda$ is treated 
as an adjustable parameter of the model.
Unfortunately, it is very difficult to perform such a quantitative 
calculation with sufficient reliability, however.
We therefore do not discuss this problem hereafter in this paper.

It should also be noted that the cutoff $\Lambda$ needs to be
fine-tuned to its critical value in order to obtain hierarchy between
the cutoff $\Lambda$ and the fermion mass $B(M_0^2)$.
The precise prediction of the cutoff and the order of fine-tuning
depend on the detail of the model parameter, however.
Actually, the scaling relation for $N_f=5$
(Fig.\ref{fig:sd6-msbar1n-5.eps}) indicates that the critical cutoff is
much larger, $\Lambda^2 \gtrsim 90|(\Q)^2|$, in $N_f=5$ model.

\section{The gap equation with use of the effective coupling} 

We have so far investigated the dynamical chiral symmetry breaking and
the phase structure in the vector-like gauge theories with extra
dimensions. 
Especially, we found that the simplest ACDH version of the top-mode
standard model ($D=4+2$, $N_f=2$) is in its chiral symmetric phase,
indicating that the 
simplest ACDH scenario does not work properly as a model to explain
the mass of the weak gauge bosons.

Our results, however, rely on our bold assumption, i.e., the non-perturbative
existence of the UV-FP\@.
If the gauge coupling becomes stronger than our estimate of the UV-FP, 
there is a chance to obtain dynamical electroweak symmetry breaking
even within the simplest ACDH model.
Moreover, there is no justification to identify the renormalization
scale $\mu^2$ of the $\overline{\rm MS}$ scheme with the gauge boson
momentum $z\equiv -q^2$ beyond the leading order in the improved
ladder approximation. 

It is therefore worth analyzing the gap equation with use of different 
definition of the gauge coupling.
Hereafter, we will investigate numerically the gap equation with use
of the effective gauge coupling defined in Ref.~\cite{TMSM_ED}.
The effective coupling is closely related to the gauge boson
propagator and its momentum.

The effective gauge coupling $g_{\rm eff}$ in the truncated KK
effective theory on the 3-brane is given by
\begin{eqnarray}
\lefteqn{
  \dfrac{-i}{g_{\rm eff}^2(-q^2)} D_{\mu\nu}^{-1}(q) \equiv
} \nonumber\\
& &  \dfrac{-i}{g_0^2} D_{(0)\mu\nu}^{-1}(q)
  - ( q^2 g_{\mu\nu} - q_\mu q_\nu )\Pi(q^2),
\label{eq:eff1}
\end{eqnarray}
with $g_0$ being bare gauge coupling of the truncated KK effective
theory. 
$D_{\mu\nu}$ and $D_{(0)\mu\nu}$ are normalized as
\begin{subequations}
\begin{eqnarray}
&&\hspace*{-6mm}
  D_{\mu\nu}(q) = \dfrac{-i}{q^2}\left(
    g_{\mu\nu}-(1-\xi(q^2))\dfrac{q_\mu q_\nu}{q^2}
  \right), \\
&&\hspace*{-10mm}
  D_{(0)\mu\nu}(q) = \dfrac{-i}{q^2}\left(
   g_{\mu\nu}-(1-\xi_0(q^2))\dfrac{q_\mu q_\nu}{q^2}
  \right).
\end{eqnarray}
\end{subequations}
Precise definitions of other notations are given in Ref.~\cite{TMSM_ED} and 
the vacuum polarization function 
$( q^2 g_{\mu\nu} - q_\mu q_\nu )\Pi(q^2)$ is evaluated 
using the background gauge fixing method so as to keep manifest gauge
invariance.
The vacuum polarization function $\Pi$ is the sum of
loop contributions of each KK-modes.
It includes not only logarithmically divergent contributions but
also finite loop corrections.

Summing up the KK-mode contributions upto $m_{KK} \le \Lambda_g$, we
obtain a relation between effective and $\overline{\rm MS}$
couplings, 
\begin{widetext}
\begin{equation}
  \dfrac{1}{\hat g_{\rm eff}^2(z)} = 
    \dfrac{\mu^2}{z}\left(
       \dfrac{1}{\hat g_{\overline{\rm MS}}^2(\mu)}
      -\dfrac{1}{g_*^2}
    \right)
   +\dfrac{1}{(4\pi)^3}\left(
     K_g(-z, \Lambda_g^2) + K_b(-z, \Lambda_g^2) + K_f(-z, \Lambda_g^2)
    \right), 
\label{eq:eff2}
\end{equation}
at one-loop level in $D=4+2$ dimensions.
Definitions of $K_g(q^2,\Lambda_g^2)$, $K_b(q^2,\Lambda_g^2)$
and $K_f(q^2,\Lambda_g^2) $ are given in Appendix~\ref{eff_coupl}. 
We also defined dimensionless bulk gauge couplings 
$\hat g_{\rm eff}^2(z)$ and $\hat g_{\overline{\rm MS}}(\mu)$ in a
similar manner to Eq.(\ref{eq:dimless}),
\begin{equation}
    \hat g^2_{\rm eff}(z) 
    = \dfrac{(2\pi R\sqrt{z})^2}{n} g^2_{\rm eff}(z), \qquad
    \hat g^2_{\overline{\rm MS}}(\mu) 
    = \dfrac{(2\pi R \mu )^2}{n} g^2_{\overline{\rm MS}}(\mu).
\end{equation}
Plugging the solution of the $\overline{\rm MS}$ renormalization group 
equation Eq.(\ref{eq:msbar6}) into Eq.(\ref{eq:eff2}), we can confirm
the renormalization scale independence of the effective coupling,
\begin{equation}
  \dfrac{1}{\hat g_{\rm eff}^2(z)} = 
    - \dfrac{1}{g_*^2} \dfrac{(\Q)^2}{z} 
   +\dfrac{1}{(4\pi)^3}\left(
     K_g(-z, \Lambda_g^2) + K_b(-z, \Lambda_g^2) + K_f(-z,\Lambda_g^2)
    \right).
\label{eq:g_eff_K}
\end{equation}
The $\overline{\rm MS}$ gauge coupling in the ACDH scenario is in the
weakly interacting region $(\Q)^2<0$.

It should be emphasized that the effective coupling Eq.(\ref{eq:eff2}) 
depends on the choice of cutoff $\Lambda_g$, no matter how large
it is. 
This behavior implies the violation of the decoupling theorem.
The low energy ($\ll \Lambda_g$) predictions are sensitive to the
physics at the cutoff scale.
There is no UV-FP in the usual sense due to the violation of the
decoupling theorem, although there still exists an upper bound of 
$\hat g_{\rm eff}$ if $(\Q)^2 \le 0$, 
\begin{equation}
  \hat g_{\rm eff}^2(z) \le 
    \left. \hat g_{\rm eff}^2(z=\Lambda_g^2) \right|_{(\Q)^2=0}
    = \dfrac{(4\pi)^3}{K},
\label{eq:max_geff}
\end{equation}
with 
\begin{equation}
  K \equiv 
  K_g(-\Lambda_g^2, \Lambda_g^2) + K_b(-\Lambda_g^2, \Lambda_g^2) 
  + K_f(-\Lambda_g^2, \Lambda_g^2), 
\end{equation}
where we identified $\Lambda_g$ with the cutoff for the gauge boson
propagator.
The factor $K$ is evaluated in $SU(N)$ gauge theory with $N_f$
flavors,
\begin{equation}
 K \equiv N \left( -\frac{88}{45}
                     +\frac{10\sqrt{5}}{3}\arctanh\frac{1}{\sqrt{5}}\right)
 - \frac{32}{45}N_f \simeq 1.63 N-0.71 N_f .
\end{equation}
We note that the upper bound of $\hat g_{\rm eff}^2$ is roughly twice
larger than the corresponding $\overline{\rm MS}$ UV-FP\@.

\end{widetext} 

%
%
\begin{figure}[t]
  \begin{center}
   \resizebox{0.45\textwidth}{!}{\includegraphics{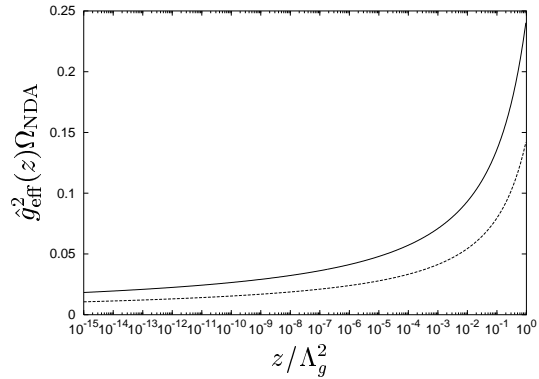}}
    \caption{The typical flow of the effective coupling 
             $\hat g_{\rm eff}^2$ defined by Eq.~(\ref{eq:g_eff_K})
             with $(\Q)^2=0$. 
             The dashed and solid lines represents the graphs for 
             $N_f=2,4$, respectively. 
             In both graphs, we took $N=3$, and
             $(\Lambda_g R)^2=4 \times 10^{10}$. \label{fig:geff}}
  \end{center}
\end{figure}

\begin{figure}[t]
  \begin{center}
   \resizebox{0.45\textwidth}{!}{\includegraphics{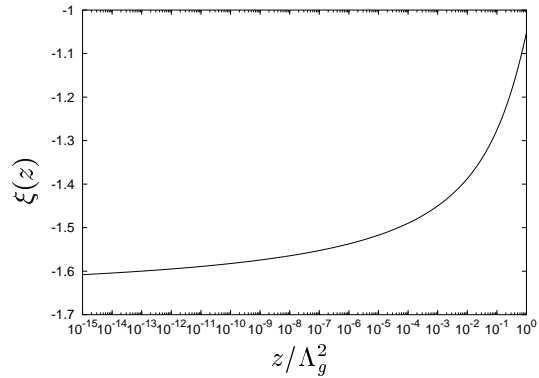}}
    \caption{The typical behavior of the NLG $\xi(z)$ 
             for the effective coupling defined by
             Eq.~(\ref{eq:g_eff_K}) with $(\Q)^2=0$. 
             We took $N=3$, $(\Lambda_g R)^2= 4 \times 10^{10}$, 
             and $N_f=4$. 
             The graph for $N_f=2$ almost overlaps with the above one. 
             The NLG approaches $\xi=-5/3$ 
             in the limit of $z \to 0$. 
             \label{xi}}
  \end{center}
\end{figure}

Fig.~\ref{fig:geff} shows typical behavior of the effective coupling 
$\hat g_{\rm eff}^2$ with $(\Q)^2=0$.
Substituting the effective coupling shown in Fig.~\ref{fig:geff} into
the formula of the non-local gauge fixing Eq.(\ref{eq:non-local2}), we
obtain the corresponding behaviors of the non-local gauge fixing
function $\xi(z)$. (See Fig.~\ref{xi}.)

In order to solve the gap equation,
we first recall Eq.(\ref{eq:wdef}), the relation between the gauge
boson momentum $z$ and the fermion momenta $x,y$ in the gap equation.
The gauge boson momentum $z$ reaches its maximum $4\Lambda^2$ when
$x=y=\Lambda^2$ and $\cos\theta=-1$ in Eq.(\ref{eq:wdef}). 
The cutoff of the gauge boson momentum $\Lambda_g^2$ thus needs to
satisfy  
\begin{equation}
  \Lambda_g^2 \ge 4 \Lambda^2.
\end{equation}
Hereafter we simply assume $\Lambda_g^2 = 4 \Lambda^2$
unless noted otherwise. 

It is difficult fully to take into account the effect of the
compactification scale $R^{-1}$.
We introduce an infrared cutoff $M_0^2 \sim R^{-2}$ in the gap
equation and neglect $R^{-1}$ sensitive infrared behaviors in the
following analysis instead.
We find the dynamical chiral symmetry breaking is insensitive to $M_0$
for sufficiently large $\Lambda$ anyway.

The minimal ACDH scenario corresponds to $SU(3)$ gauge theory in the
$D=4+2$ dimensional space-time with $N_f=2$ and $(\Q)^2<0$.
The effective gauge coupling of the ACDH scenario $(\Q)^2<0$ is always
weaker than the case with $(\Q)^2=0$.
It is therefore sufficient to investigate the case $(\Q)^2=0$ for the 
determination of the condition of the bulk chiral symmetry breaking.
The aim of our numerical analysis is then to find the critical $N_f$,
above which the dynamical chiral symmetry breaking takes place in the
bulk with $(\Q)^2=0$.
We take the chiral limit $m_0=0$ in the following analyses.

\begin{figure*}[t]
  \begin{center}
    \resizebox{0.45\textwidth}{!}{\includegraphics{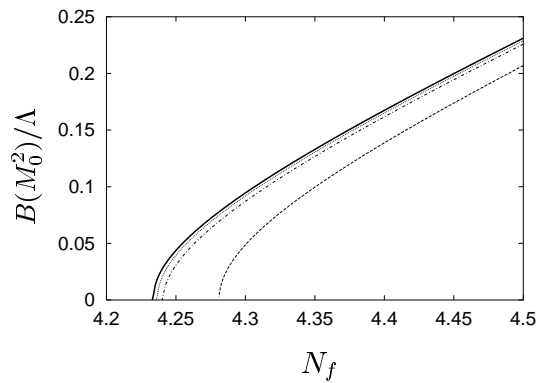}}
    \caption{The scaling behavior for the dynamical mass with 
             the effective coupling $\hat g_{\rm eff}^2$. 
             The lines from right to left are graphs for 
             $(\Lambda/M_0)^2=10^2, 10^3, 10^5, 10^{10}$ with $N=3$ and 
             $(\Lambda_g/\Lambda)^2=4$, respectively. 
             \label{fig:criNf}}
  \end{center}
\end{figure*}

It is now easy to perform numerical analysis of the gap equation by
using the recursion method \cite{TMSM_ED}.
Hereafter, we formally admit $N_f$ to take non-integer (real) value
and evaluate the scaling behavior of $B(M_0^2)$ as a function of $N_f$.
For $SU(3)$ gauge theory in $D=4+2$ dimensions, 
we obtain the scaling behavior shown in Fig.\ref{fig:criNf}.
The dynamical chiral symmetry breaking takes place for 
\begin{equation}
 N_f > N_f^{\rm crit} = 4.23  \quad \mbox{with } (\Lambda_g/\Lambda)^2=4 
 \label{Ncri_geff}
\end{equation}
in this model.

%
%

For $\Lambda_g^2 > 4\Lambda^2$,
critical $N_f$ tends to be lager than Eq.~(\ref{Ncri_geff}). 
For instance, we obtain
\begin{equation}
 N_f^{\rm crit} = 4.62  \quad \mbox{with } (\Lambda_g/\Lambda)^2=10. 
\end{equation}
The physics behind this result is obviously understood if we note 
Eq.(\ref{eq:max_geff}). 
The effective coupling does not reach its maximum value in the gap
equation for $\Lambda_g^2 > 4\Lambda^2 \geq z$.
We thus conclude that Eq.(\ref{Ncri_geff}) is very conservative
estimate and that 
the simplest version of the ACDH model with $N_f=2$ 
does not work even with the effective coupling.

%
%

Noting the upper bound of $\hat g_{\rm eff}^2$ is approximately twice
larger than the UV-FP of $\hat g_{\overline{\rm MS}}^2$,
it is somewhat surprising to find that $N_f^{\rm crit}$ with the
effective coupling Eq.(\ref{Ncri_geff}) is rather close to $N_f^{\rm crit}$ 
with the $\overline{\rm MS}$ coupling Eq.(\ref{eq:Ncri_MS}).
It should be emphasized that Fig.\ref{fig:geff} shows the
effective coupling $\hat g_{\rm eff}^2$ is close to its maximum value
only when the gauge boson momentum $z$ is sufficiently close to
$\Lambda_g^2$, however.
Unlike the corresponding UV-FP of the $\overline{\rm MS}$ coupling, 
the effective coupling is well below its maximum value in the wide
region of momentum space.

Actually, similar situation was also found in the analysis of
four-dimensional QED with including vacuum polarization
effects~\cite{QED4vac}.
In the case of $N_f=1$ QED${}_4$, the dynamical chiral symmetry
breaking takes place only when the coupling at the cutoff exceeds the
critical value 
\begin{equation}
 \alpha_\Lambda > 1.95 , \quad (N_f=1) , 
\end{equation}
which is about twice larger than the quenched one ($\alpha_c=\pi/3$). 

\begin{figure}[hbt]
  \begin{center}
   \resizebox{0.45\textwidth}{!}{\includegraphics{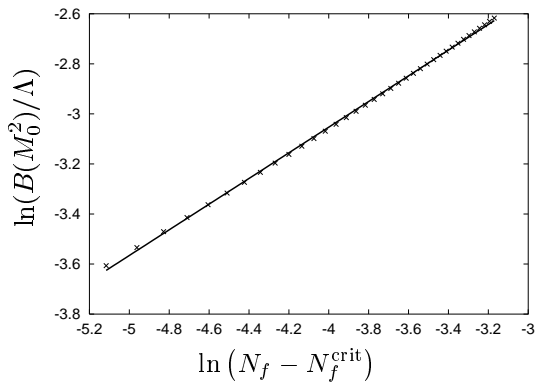}}
    \caption{The log-log plot of the scaling relation Fig.~\ref{fig:criNf}.
             We used $(\Lambda/M_0)^2=10^{10}$ and $(\Lambda_g/\Lambda)^2=4$.
             The bold line represents the line 
             obtained through the least-squares method for the data set of 
             $(N_f-N_f^{\rm crit})/N_f^{\rm crit} < 0.01$ 
             with $N_f^{\rm crit}=4.23$. 
             We also plot numerical data with crossed points. 
             \label{fig:power}}
  \end{center}
\end{figure}

We finally make a brief comment on the scaling relation.
Unlike the essential-singularity type scaling Eq.(\ref{eq:sol-sd1})
found in the analysis with the $\overline{\rm MS}$ coupling, 
the scaling behavior of Fig.~\ref{fig:criNf} seems like the mean-field
type scaling.
In order to confirm the mean-field type scaling, we perform a fit of
the scaling behavior assuming power-low scaling,
\begin{equation}
 B(M_0^2) \propto  
 \Lambda \left( N_f - N_f^{\rm crit} \right)^{\gamma},
\end{equation}
in Fig.\ref{fig:power}.
We find the best fit value $\gamma$ is given by 
\begin{equation}
 \gamma = 0.51
\end{equation}
for the data set with
\begin{equation}
  \frac{N_f - N_f^{\rm crit}}{N_f^{\rm crit}} \leq 0.01.
\end{equation}
This result is consistent with the mean-field type scaling
$\gamma=1/2$.
In contrast to the case with the essential-singularity type scaling, 
the cutoff $\Lambda$ needs to be small enough to keep the dynamical 
mass small even when $N_f$ is sufficiently close to $N_f^{\rm crit}$
with the mean-field type scaling.
  
%
%
\printfigures

\section{Summary and Discussions}
\label{summary}

In this paper, we have systematically
studied the bulk
dynamical chiral symmetry breaking in vector-like gauge theories with
extra dimensions and revealed a new phase structure of such theories.
Extending our previous study based on the gap equation (the SD equation within
the improved ladder approximation),
we adopted in the present study
the non-local gauge fixing method 
in order to keep the
$\fsl{p}$ part of the fermion propagator to be trivial, i.e.,
$A(-p^2)\equiv 1$, which is thus consistent with the Ward-Takahashi
identity and the bare vertex approximation in the ladder SD equation. 

The one-loop analysis of the $\overline{\rm MS}$ beta function suggests
the existence of a 
ultra-violet fixed-point (UV-FP) $g_*$ in the truncated KK effective
theory of a non-Abelian gauge theory with compactified extra dimensions. 
The existence of such a UV-FP may open interesting
possibilities in the model building of the high-energy particle
theory. 
The 
top mode standard model (TMSM) 
scenarios in extra dimensions, for example,
are affected significantly by the existence of such a UV-FP\@.
It is therefore interesting to investigate consequences of a 
UV-FP in the bulk field theories.
We therefore first analyzed the gap equation with the gauge coupling
both on and off the UV-FP, assuming the qualitative structure of the
UV-FP is unchanged beyond the one-loop approximation.
We found
that the critical UV-FP gauge
coupling is $D/4$ times larger than our previous calculation in the
Landau gauge. 
The result was then converted to {\em critical number of flavors}
$N_f^{\rm crit}$.
For $N_f > N_f^{\rm crit}$ the dynamical chiral symmetry breaking
takes place not only in the ``strong-coupling phase'' $\hat g > g_*$,
but also in the  ``weak-coupling phase'' $\hat g < g_*$ when the
cutoff is large enough.
For $N_f < N_f^{\rm crit}$, however,
the chiral symmetry remains unbroken in the ``weak-coupling phase''
$\hat g < g_*$ no matter how large the cutoff is.
We found $N_f^{\rm crit} \simeq 4.2, 1.8$ in $D=6,8$ dimensions for
the $SU(3)$ gauge theory (bulk QCD).

In a scenario with the extra dimensions (ACDH
scenario) of the TMSM, 
the gauge coupling is obviously weak and the $N_f$ needs
to be larger than the critical one in order to trigger the dynamical 
electroweak symmetry breaking.
The simplest ACDH scenario with $N_f=2$ thus does not work in $D=6$,
while there is a chance to construct a viable model in $D=8$
dimensions.
Moreover, the UV cutoff needs to be finite in order to obtain finite
top quark mass. 
Actually, once we fix the top mass, 
it is 
possible to predict the UV cutoff in
the ACDH scenario in the analysis of the gap equation, in contrast to
the original treatment of ACDH\@. 
The phenomenological analysis done by ACDH 
therefore
needs to be modified
by taking account of this fact.

On the other hand, we found a novel situation for $N_f<N_f^{\rm
crit}$, where we can formally define a continuum limit (infinite
cutoff). 
The low energy physics is controlled by the properties of the UV-FP
and it becomes insensitive to the physics around the UV cutoff.
The anomalous dimension of the fermion mass $\gamma_m$ is shown to be
large. 
This phase may be useful for model building of ``bulk
technicolor'', where the large anomalous dimension can be used as a
suppression mechanism of the excess of FCNC\@.

It should be emphasized, however, that non-perturbative existence of
UV-FP is no
more than 
an assumption at present in 
a wide class of models. 
Actually, the one-loop effective gauge coupling of the truncated KK
effective theory with $D=4+2$ dimensions is shown to have an explicit
cutoff dependence, which implies 
absence of the UV-FP in the usual
sense.
We therefore performed an analysis of the gap equation by using the
effective coupling for $D=4+2$ dimensions.
We found that 
there also exists a $N_f^{\rm crit}$, notwithstanding absence
of UV-FP in the usual sense.
Although the effective coupling at the UV cutoff is much larger
than that of 
$\overline{\rm MS}$, we found that 
the $N_f^{\rm crit}$ in this scheme actually
is very close to the $\overline{\rm MS}$ one.
The simplest version of ACDH scenario with $N_f=2$ is therefore quite
unlikely to work in the $D=4+2$ dimensions.

Many issues remain unsolved and need further study, however.
For a example, the existence of a non-trivial UV-FP is yet to be
proved. 
It should be investigated more definitely in future whether
the non-trivial UV-FP really exists or not.

The uncertainties coming from compactification
sensitive infrared (IR) region are also important.
In the present paper, this effect was only roughly estimated by
introducing the IR cutoff $M_0^2 \sim R^{-2}$ in the gap equation. 
It turned out that 
the effects  tend to increase the critical coupling or $N_f^{\rm crit}$. 
Hence there is a possibility that the simplest version of the ACDH scenario 
may not work even for $D=4+4$. 
We need to invent more sophisticated way to take into account the
compactification effect particularly for the case $B_0 \ll R^{-1}$.

Finally, the results presented in this paper provide basic tools for the
particle model building with the dynamical chiral symmetry breaking in
extra dimensions. 
We need to construct concrete and viable models such as top-condensate or
technicolor in the bulk by using these tools.
More quantitative studies on these models will be dealt in a
separated publication.

\begin{acknowledgments}
This work is supported by Grant-in-Aid for
Scientific Research (B) $\#$11695030 (K.Y., V.G.),  
the JSPS Research Fellowships for Young Scientists $\#$01170 (M.H.),
and partially by the grant SCOPES project $7$ IP $062607$ of the Swiss
NSF (V.~G.). 
V.G. wishes to acknowledge JSPS for financial support. 
\end{acknowledgments}

\appendix

\section{The Effective potential for local composite field}
\label{app:eff_pot} 

We expect appearance of various composite $\bar\psi\psi$ bound
states in the gauge theories with extra dimensions.
The $\sigma$, chiral partner of the Nambu-Goldstone boson, is
particularly 
important among them, because it corresponds to the Higgs boson in the 
models of the dynamical electroweak symmetry breaking.
However, the CJT potential discussed in section \ref{sec:fp_solution} 
is a
functional of the mass function $B(x)$, and is not directly related to
these bound states.
In addition, it is shown that the CJT is not bounded from below.
It is therefore not perfectly appropriate to study
the stability of the vacuum by using the CJT potential.

In this Appendix, we thus discuss yet another effective
potential $V(\sigma)$, which is a function of the local
composite field $\sigma \sim \bar\psi\psi$ and is connected to the
dynamical properties of the $\sigma$ boson more closely. 

We consider the effective action $\Gamma[\sigma]$,
\begin{widetext}
\begin{equation}
  \Gamma[\sigma] \equiv W[J] - \int d^D x J \sigma,
\end{equation}
with 
\begin{equation}
  W[J] \equiv \frac{1}{i} \ln \int [d\psi d\bar\psi] [{\rm gauge}]
    \exp \left( i\int d^D x ({\cal L} + J \bar\psi\psi) \right),
  \qquad
  \sigma \equiv \dfrac{\partial W[J]}{\partial J}.
\end{equation}
\end{widetext} 
The corresponding effective potential can be obtained by taking
the coordinate independent part of this effective action.

In the following, we briefly outline the derivation of the effective
potential  $V(\sigma)$ based on the method of Ref.~\cite{eff_pot} 
(see also Ref.~\cite{MY97}).

For a constant source term $J$, the partition function $W[J]$ is
obtained as
\begin{equation}
  W[J] = \int dJ \dfrac{\partial W[J]}{\partial J} \int d^D x.
\end{equation}
Noting 
\begin{equation}
  \dfrac{\partial W[J]}{\partial J} = \VEV{\bar\psi\psi}_J  = \sigma,
\end{equation}
we find the effective potential is given by
\begin{equation}
  V(\sigma) = J \sigma - \int^J dJ \sigma \left(=\int^\sigma d\sigma J\right),
\label{eq:localpot1}
\end{equation}
where $J$ should be regarded as a function of $\sigma$.

The effect of the constant source $J$ can be obtained by replacing the
bare mass $m_0$ in Eqs.(\ref{eq:gap2}), (\ref{eq:uv-bc}),
(\ref{eq:uv-bc2}) and (\ref{eq:crit-uv-bc}),
\begin{equation}
  m_0 \rightarrow m_0 - J.
\label{eq:replace}
\end{equation}
We thus find
\begin{widetext}
\begin{equation}
  J = m_0 - 
  \frac{1}{2}(1+\tilde \nu) \tilde c_0
  B_0\left(\dfrac{\Lambda^2}{B_0^2}\right)^{-\omega(1-\tilde\nu)},
\label{eq:Jexpr-1}
\end{equation}
for $\kappa_D < \kappa_D^{\rm crit}$ and
\begin{equation}
  J = m_0 -  
  \sqrt{1+\nu^2} |c_0| B_0 \left(\dfrac{\Lambda^2}{B_0^2}\right)^{-\omega}
         \sin\left[\theta + \omega\nu \ln \frac{\Lambda^2}{B_0^2}
                          +\tan^{-1}\nu 
             \right],
\label{eq:Jexpr-2}
\end{equation}
for $\kappa_D > \kappa_D^{\rm crit}$.
The relation between the mass function $B(x)$ and $\sigma$ is given by
\begin{equation}
  \sigma = \VEV{\bar\psi\psi}_J
    = -\eta N N_f \Omega_{\rm NDA} \int_0^{\Lambda^2} dx x^{D/2-1} 
       \dfrac{B(x)}{x+B^2},
\end{equation}
which leads to 
\begin{equation}
  \sigma = \VEV{\bar\psi\psi}_J 
    = \eta N N_f \Omega_{\rm NDA} 
      \dfrac{\kappa_D^{\rm crit}}{\omega^2 \kappa_D}  \Lambda^{D} \left. 
      \frac{d B(x)}{dx}\right|_{x=\Lambda^2}, 
\label{eq:sigmaexpr}
\end{equation}
where we have used the gap equation Eq.(\ref{eq:gap2}). 
The chiral condensate $\sigma$ can be expressed in terms
of $B_0$,
\begin{equation}
  \sigma(B_0) = - \tilde c_0 \eta N N_f \Omega_{\rm NDA}
  \dfrac{(1-\tilde{\nu})\kappa_D^{\rm crit}}{\omega \kappa_D}
  \Lambda^{D-1}
  \left(\dfrac{\Lambda}{B_0}\right)^{-2\omega(1-\tilde \nu)-1},
\label{eq:Bcutoff1}
\end{equation}
for $\kappa_D < \kappa_D^{\rm crit}$ and
\begin{equation}
  \sigma(B_0) = - 
  \dfrac{2| c_0 | \eta N N_f \Omega_{\rm NDA}}{\omega\sqrt{1+\nu^2}}
  \Lambda^{2\omega} B_0^{2\omega+1}
  \sin\left[\theta + \omega\nu\ln\dfrac{\Lambda^2}{B_0^2}-\tan^{-1}\nu\right],
  \label{sigma}
\end{equation}
for $\kappa_D > \kappa_D^{\rm crit}$.

We are now ready to evaluate the effective potential
Eq.(\ref{eq:localpot1}).
The $J$ (or $\sigma$) integral in Eq.(\ref{eq:localpot1}) can be performed 
by using
\begin{equation}
  dJ = d B_0  \dfrac{dJ}{dB_0}, \quad 
 \left( \mbox{or} \quad d\sigma = d B_0 \dfrac{d\sigma}{d B_0} \right).
\end{equation}

Combining Eqs.(\ref{eq:localpot1}), (\ref{eq:Jexpr-1}),
(\ref{eq:sigmaexpr}) and (\ref{eq:Bcutoff1}), we finally obtain
the effective potential in the subcritical region  $\kappa_D <
\kappa_D^{\rm crit}$  
\begin{equation}
  V(\sigma) 
  = \dfrac{1}{D-2} \eta N N_f \Omega_{\rm NDA} 
    \tilde c_0^2 \Lambda^D
    \left(\dfrac{B_0^2}{\Lambda^2}\right)^{1+2\omega(1-\tilde \nu)},
\label{eq:localpot2}
\end{equation}
in the chiral limit $m_0=0$.
The effective potential in the super-critical region can be obtained
in a similar manner,
\begin{equation}
  V(\sigma) = \dfrac{2}{D-2} \eta N N_f \Omega_{\rm NDA} 
  |c_0|^2 B_0^D 
  \left[
    - \cos\left(
        2\theta + 2\omega\nu \ln\dfrac{\Lambda^2}{B_0^2}
      \right)
    + A 
  \right], \label{eq:localpot3}
\end{equation}
with
\begin{displaymath}
  A \equiv \dfrac{\kappa_D^{\rm crit}}{\kappa_D}
    \left( 1 - \dfrac{1+6\omega}{1+2\omega} \nu^2 \right).
\end{displaymath}
We regard here $B_0$ as an function of $\sigma$ defined implicitly in
Eq.(\ref{eq:sigmaexpr}). 

\end{widetext} 

It is now straightforward to find the stationary points of the 
effective potential.
The stationary condition $dV/d\sigma = 0$ of
Eq.(\ref{eq:localpot2}) has only the trivial solution
$B_0=0$ for $\kappa_D < \kappa_D^{\rm crit}$, while
we find non-trivial solutions $B_0=B_0^{(n)} \ne 0$ ($n=1,2,\cdots$)
for $\kappa_D > \kappa_D^{\rm crit}$ in the stationary condition 
of Eq.(\ref{eq:localpot3}).
Here, $B_0^{(n)}$ is given by
\begin{equation}
B_0^{(n)} \equiv 
\Lambda \exp\left[
  \dfrac{-n\pi +\theta + \tan^{-1}\nu}{2\omega\nu} \right],
\end{equation}
which coincides with the solution of Eq.(\ref{eq:crit-uv-bc}) with
$m_0=0$. 

Hereafter, we concentrate on the supercritical region $\kappa_D >
\kappa_D^{\rm crit}$.
The stability of the vacua ($n=1,2,\cdots$) can be investigated by
taking the second derivative of the potential, 
\begin{equation}
\dfrac{d^2 V}{d\sigma^2} = \dfrac{dJ}{d\sigma} 
= \dfrac{dJ}{dB_0} \cdot \left(\dfrac{d\sigma}{dB_0}\right)^{-1}. \label{ddV}
\end{equation}
It is easy to show that 
the curvature of the potential at stationary point is positive 
\begin{widetext}
\begin{eqnarray}
\left.\frac{d^2V}{d\sigma^2}\right|_{\sigma=\sigma(B_0^{(n)})}
= \frac{\omega^2(1+\nu^2)}
       {2(1+3\omega-\omega\nu^2)\eta N N_f \Omega_{\rm NDA}}\Lambda^{-(D-2)}
 > 0 
\end{eqnarray}
for $\nu^2 < 3+\omega^{-1}$ irrespective of $n$.
We thus find that every stationary point is a local minimum of the
potential $V(\sigma)$. 

We need to compare vacuum energies in order to find the absolute
minimum of the potential, i.e., the true vacuum then.
The value of the potential at each $n$ is obtained as
\begin{eqnarray}
V(B_0^{(n)}) = -\frac{4}{D}
 \left(1-\frac{\kappa_D^{\rm crit}}{\kappa_D}\right)
 |c_0|^2 \eta N N_f \Omega_{\rm NDA}(B_0^{(n)})^{D} < 0,
\end{eqnarray}
which is actually consistent with the result of the CJT potential  
Eq.~(\ref{eq:vac-energy}) for small $B_0^{(n)}$. 
The $n=1$ solution, i.e., the largest fermion mass, gives the global minimum
of $V(\sigma)$.
We thus conclude that the $n=1$ solution corresponds to the most
stable vacuum.

We should comment here on the properties of the false vacua $n \ge 2$.
Although we found that the mass square of $\sigma$ is positive even in
these false vacua, it does not necessarily imply the meta-stability of
these vacua. 
Actually, in the analysis of the Bethe-Salpeter equations in the
strong coupling QED$_4$ (QED in four dimensions), it is known that
these false vacua have tachyonic mode(s) in the pseudoscalar channel
in addition to the massless Nambu-Goldstone
mode~\cite{RNC_review,DSBbook}.
The false vacua $n \ge 2$ are therefore saddle points of the effective
potential with negative curvature in the direction of the pseudoscalar
channel when pseudoscalar degrees of freedom are included in the
potential.


As we described before, we found that the every stationary point of the
potential Eq.(\ref{eq:localpot3}) is local minimum in the
$\sigma$-direction.
There is no stationary point corresponding to local maximum.
This fact perhaps may sound rather peculiar. 
Actually, it is tied with the interesting and bizarre property of the
effective potential Eq.(\ref{eq:localpot3}).
The effective potential Eq.(\ref{eq:localpot3}) is a
multi-branched and multi-valued function of $\sigma$.
We next try to grasp the shape of the effective potential more closely.

We first consider the derivatives of $\sigma$ (Eq.(\ref{sigma})) and
$J$ (Eq.(\ref{eq:Jexpr-2})) with respect to $B_0$:
\begin{equation}
\dfrac{d \sigma}{d B_0} \propto 
 \Lambda^{2\omega} B_0^{2\omega}
 \sin \left(\theta+\omega\nu\ln\frac{\Lambda^2}{B_0^2}
-\tan^{-1}\left[\frac{2(D-1)\nu}{D-(D-2)\nu^2}\right]\right), 
\label{dsdb0} 
\end{equation}
and 
\begin{equation}
\dfrac{d J}{d B_0} \propto \Lambda^{-2\omega} B_0^{2\omega}
 \sin \left(\theta+\omega\nu\ln\frac{\Lambda^2}{B_0^2}
+\tan^{-1}\left[\frac{2\nu}{D+(D-2)\nu^2}\right]\right).
\label{dJdb0}
\end{equation}
We note that $d\sigma/dB_0$ vanishes for
\begin{equation}
 \theta + \omega\nu\ln\frac{\Lambda^2}{B_0^2} 
  -\tan^{-1}\left(\frac{2(D-1)\nu}{D-(D-2)\nu^2}\right)
 = n \pi, \qquad n=1,2,\cdots   \label{maxima}
\end{equation}
while $dJ/dB_0$ remains finite for $B_0$ with Eq.(\ref{maxima}).
The second derivative of the potential Eq.(\ref{ddV}) thus diverges at
Eq.(\ref{maxima}). 
We note here, however, the first derivative of the potential 
$d V/d\sigma(= dV/dB_0 \cdot dB_0/d\sigma)$ remains finite 
at Eq.(\ref{maxima}) despite $dV/dB_0=0$, since there exists in
$dV/dB_0$ the same sine-function as in $d\sigma/dB_0$.
There should take place something very bizarre at the points
Eq.(\ref{maxima}). 
\end{widetext} 

Plotting the shape of the effective potential, 
we find that it has structure quite similar to the potential Fig.1 of
Ref.~\cite{eff_pot_dqcd}, in which the diquark condensate is studied
in high density QCD by using the local composite effective
potential. 
The potential $V(\sigma)$ is a multi-branched and multi-valued
function of $\sigma$: The points Eq.~(\ref{maxima}) are cusps and
correspond to branching points. 
The stationary point $\sigma(B_0^{(n)})$ corresponds to the local minimum 
in each branch of the effective potential. 
We also find easily from Eqs.~(\ref{maxima}) and (\ref{sigma}) 
that the branching point converges to $\sigma=0$ 
in the $n \to \infty$ limit. 
This means that the branch of the potential also shrinks into
$\sigma=0$, exhibiting a fractal structure around $\sigma=0$.

It is known that the long-range nature of interactions in scale
invariant theories also leads to other peculiar properties such as the
existence of the infinite number of resonances and the non-analyticity
of the potential at the point $\sigma(B_0=0)$ causing its fractal
structure around $\sigma(B_0=0)$~\cite{eff_pot_dqcd}. 
It should be noted, however, that the gauge theories with extra
dimensions are not scale invariant below the compactification scale
which should serve as an infrared cutoff in the SD equation for the
fermion mass function. 
Introducing such a cutoff explicitly, one can show that the SD
equation has only finite number of  solutions (see
Refs.~\cite{QED4_BSamp,RNC_review} and Appendix C of the present
paper) and the solutions with small 
dynamical masses $B_0^{(n \gg 1)}$ disappear.
Accordingly, the potential $V(\sigma)$ will have only finite number of 
branches. 
We thus expect that bizarre behavior near $\sigma=0$ does not actually
occur in models treated in this paper.

Finally, we discuss the properties of the effective potential for
sufficiently large cutoff $\Lambda \gg B_0^{(1)}$.
For such a purpose, we take an infinite cutoff limit 
$\Lambda \rightarrow \infty$ with $B_0^{(1)}$ being fixed by formally
adjusting $\nu \equiv \sqrt{\kappa_D/\kappa_D^{\rm crit}-1}$.
The anomalous dimension of the fermion mass is found to be 
$\gamma_m=2\omega$ in Ref.\cite{TMSM_ED} in such a formal limit.
We thus define a ``renormalized'' operator,
\begin{equation}
  (\bar\psi\psi)_R \equiv Z_m (\bar\psi\psi), \quad
  Z_m \propto \left(\dfrac{\mu}{\Lambda}\right)^{2\omega}
\end{equation}
and 
\begin{equation}
  \sigma_R \equiv \VEV{(\bar\psi\psi)_R}_J.
\end{equation}
Taking the formal $\Lambda\rightarrow\infty$ limit of Eq.(\ref{sigma})
as described before, it is easy to obtain
\begin{equation}
  \sigma_R \propto
    \mu^{2\omega} B_0^{D/2} \left(
      \dfrac{1}{\omega} + \ln \dfrac{B_0}{B_0^{(1)}}
    \right).
\label{eq:sigma_R}
\end{equation}
As expected from the argument of the anomalous dimension, $\sigma_R$
remains finite even in this formal $\Lambda\rightarrow\infty$ limit.

Eq.(\ref{eq:sigma_R}) can be used to define $B_0$ as a function of
$\sigma_R$.
We are thus able to rewrite the effective potential
Eq.(\ref{eq:localpot3}) as a function of the renormalized field,
$V_R(\sigma_R)$. 
In the formal $\Lambda\rightarrow \infty$ limit, we obtain
\begin{widetext}
\begin{equation}
  V_R(\sigma_R)
   = \eta N N_f \Omega_{\rm NDA}
     \dfrac{[\Gamma(D/2)]^2}{[\Gamma(1+\omega)]^4}
     B_0^D \left[
       \omega\left(\ln\dfrac{B_0}{B_0^{(1)}}\right)^2
      +\ln\dfrac{B_0}{B_0^{(1)}} - \frac{1}{D}
     \right],
\label{eq:effpot_R}
\end{equation}
\end{widetext}
with $B_0=B_0(\sigma_R)$ being the function of $\sigma_R$ determined
implicitly from Eq.(\ref{eq:sigma_R}).

It is somewhat surprising to find such a finite expression of the
effective potential Eq.(\ref{eq:effpot_R}) in the $\Lambda\rightarrow
\infty$ limit in the non-renormalizable higher-dimensional gauge theories.
This property is actually related to the approximate scale invariance,
i.e., the existence of the UV-FP\@.


\section{Conversion to the Schr\"{o}dinger-like equation}
\label{schro}

We discuss the dynamical mass generation in the bulk  
from a little bit different point of view.
The SD eq.~(\ref{eq:gap2}) for the mass function 
without the bare mass term $m_0$ 
can be rewritten in a form of 
the Schr\"{o}dinger-like equation~\cite{GSSW}, in which 
the \DxSB~ takes place when a ``bound state'' exists.
The subject whether the bulk fermion condenses or not is, thus, 
reduced to the ``bound state problem'' in the quantum mechanics.

Let us start with introducing the ``wave function''
\begin{equation}
 \psi(u) \equiv \int \frac{d^D q_E}{(2\pi)^D}
 \frac{e^{i q_E \cdot u}B(q_E^2)}{q_E^2+m^2}.
 \label{wave1}
\end{equation}
The Fourier transform of the mass function $B$ is then given by
\begin{equation}
 \int \frac{d^D q_E}{(2\pi)^D} e^{i q_E \cdot u}\;B(q_E^2) =
 (-\triangle_u+m^2) \psi(u) . \label{wave2}
\end{equation}
We next consider a linearized version of the ladder SD equation,
\begin{equation}
  B(p_E^2) = C_F \int \dfrac{d^D q_E}{(2\pi)^D}
  \dfrac{B(q_E^2)}{q_E^2 + m^2} \dfrac{(D-1+\xi)
  g_*^2}{(p_E-q_E)^{2(D/2-1)}},
\label{eq:app_sdeq}
\end{equation}
where the coupling $g_D^2$ is replaced by the running one
$g_*^2/(p_E-q_E)^{2(D/2-2)}$ and we take $m=B(0)$.
It is straightforward to show that the Fourier transform of
Eq.(\ref{eq:app_sdeq}) is formally given by
\begin{eqnarray}
 \int\!\! \frac{d^D q_E}{(2\pi)^D} e^{i q_E \cdot u}\;B(q_E^2) 
&=& - V(u) \psi(u), \label{wave3}
\end{eqnarray}
where the ``potential'' $V(u)$ is defined by
\begin{equation}
 V(u) \equiv -(D-1+\xi_D) C_F g^2_* \int \frac{d^D p_E}{(2\pi)^D} 
 \frac{e^{i p_E \cdot u}}{[p_E^2]^{D/2-1}} \label{pot}
\end{equation}
and the gauge fixing parameter $\xi$ is taken to the value of
Eq.~(\ref{eq:nlg1}),
\begin{equation}
  \xi_D \equiv -\frac{(D-1)(D-4)}{D} .
\end{equation}
Here, the momentum shift invariance ($p_E \to p_E-q_E$) is 
assumed~\footnote{It should be noted, however, the UV-cutoff $\Lambda$
in the SD equation violates the momentum shift invariance.
The analysis of the Schr\"{o}dinger-like equation can thus be regarded
as an analysis of the SD equation with a different choice of the
UV-cutoff procedure.
}.
Eqs.~(\ref{wave2}) and (\ref{wave3}) then lead to the Schr\"{o}dinger-like 
equation,
\begin{equation}
 H \psi(u) = E \psi(u), \label{schro1}
\end{equation}
with
\begin{equation}
H \equiv -\triangle_u + V(u), \quad E \equiv -m^2.
\end{equation}
Namely, non-trivial solutions of the SD equation 
correspond to bound states ($E<0$) in the Schr\"{o}dinger-like equation. 

In order to solve the Schr\"{o}dinger-like equation (\ref{schro1}), 
we rewrite it in the spherical coordinate. 
The ``wave function'' is decomposed as 
\begin{equation}
\psi(u) \equiv r^{-\frac{D-1}{2}}R(r)X(\phi_i),  \quad r=|u|,
\end{equation}
where $R(r)$ and $X(\phi_i)$ denote the radial function and an analogue of 
the spherical surface harmonics in $D$-dimensions (the Gegenbauer function),
respectively. 
For a ``S-wave wave function'' with $X(\phi_i) \equiv 1$,
the $D$-dimensional Laplacian can be written as 
\begin{eqnarray}
\lefteqn{
 \triangle_u \psi_{\rm S}(u) =
} \nonumber\\
  & &
 \frac{1}{r^{D-1}}\frac{\partial }{\partial r}\left[r^{D-1}
   \frac{\partial}{\partial r}\left(r^{-\frac{D-1}{2}}R(r)\right)\right].
\end{eqnarray}
It is straightforward to show that Eq.~(\ref{schro1}) leads to 
\begin{equation}
\left[\,-\frac{\partial^2 }{\partial r^2} + V_{\rm eff}(r)\,\right] R(r) = 
 E R(r)
\end{equation}
with
\begin{equation}
 V_{\rm eff}(r) \equiv V(r) + \frac{(D-1)(D-3)}{4}\frac{1}{r^2}.
\end{equation}
We find that an additional ``{\it positive centrifugal potential}'' 
appears from the kinetic term in the case of $D > 3$ 
even if we consider the S-wave solution. 

The ``potential'' $V(r)$ in Eq.~(\ref{pot}) 
also has the same power of $r$ 
(an {\it attractive inverse square} ``potential''): 
\begin{equation}
V(r)=-\frac{(D-2)^2}{4}\frac{\kappa_D}{\kappa_D^{\rm crit}}\frac{1}{r^2}.
\end{equation}
The competition between the ``{\it repulsive} centrifugal
potential'' and  
the ``{\it attractive} inverse square potential'' thus
determines the dynamical symmetry breaking.

The bound state spectrum with an inverse square potential 
can be found in various textbooks of quantum
mechanics~\cite{morse_feshbach}.
The equation
\begin{equation}
 \frac{\partial^2 R(r)}{\partial r^2} 
 + \left[\, \epsilon + \frac{\alpha}{r^2}\,\right] R(r) = 0, 
\end{equation}
has an infinite number of bound state solutions only when $\alpha > 1/4$. 
In the present case, the parameters $\epsilon$ and $\alpha$ are given by
\begin{equation}
\epsilon = -m^2, \quad 
\alpha = \frac{(D-2)^2}{4}\left(\frac{\kappa_D}{\kappa_D^{\rm crit}}-1\right)
         +\frac{1}{4}. 
\end{equation}
The bound states of the Schr\"{o}dinger-like equation exist
if and only if $\kappa_D > \kappa_D^{\rm crit}$. 
The analysis of the Schr\"{o}dinger-like equation gives the critical
point $\kappa_D^{\rm crit}$, which coincides with the value in the
SD equation.  

We note here that the size of the ``repulsive centrifugal potential''
becomes significant for $D\gg 4$.
This is the very reason why we obtain $\kappa_D^{\rm crit}$ larger
than the NDA estimate.

We next comment on the case with non-running $g_D$~\cite{GSSW}. 
In this case, the ``potential'' $V(r)$ is given by
\begin{eqnarray}
 V(r) &\equiv& -(D-1)g_D^2 \int \frac{d^D p_E}{(2\pi)^D} 
 \frac{e^{i p_E \cdot u}}{p_E^2} \nonumber \\
&=& - (D-1)\frac{\Gamma(D/2-1)}{4\pi^{D/2}}\frac{g_D^2}{r^{D-2}} 
\end{eqnarray}
in the Landau gauge ($\xi=0$). 
When the potential behaves as $-1/r^s$ ($0 < s < 2$) for sufficiently large $r$, 
the spectrum contains a countably infinite number of 
bound states~\cite{hilbert}. 
The dynamical symmetry breaking thus occurs 
for any value of the gauge coupling in $2< D <4$~\cite{GSSW}. 
However, it is not true in $D>4$. 
There is a critical point and the scaling relation is power-like in 
the numerical analysis for $D=5,6$~\cite{KN89}. 

\section{Effects of IR cutoff in the gap equation at the fixed point}
\label{ir-cutoff}
 
In this Appendix we solve the linearized equation (\ref{eq:diff2})
in the presence of the IR cutoff $M_0\sim R^{-1}$.

The integral gap equation is written as the differential one 
\begin{widetext}
\begin{eqnarray}
\dfrac{d^2B(x)}{dx^2}+\frac{2\omega+1}{x}\frac{dB(x)}{dx}+\omega^2(1+\nu^2)
\dfrac{B(x)}{x(x+B_0^2)}=0,
\label{linearized:eq}
\end{eqnarray}
with two (infrared and ultraviolet) boundary conditions
\begin{eqnarray}
&&x^{2\omega+1}\left.\frac{d}{dx}B(x)\right|_{x=M_0^2} = 0,\quad \mbox{(IR-BC)}, 
\label{IRBC} \\  
&&   \left.\left(1+\frac{x}{2\omega}\frac{d}{dx}\right)B(x)\right|_{x 
=\Lambda^2}= m_0,\quad \mbox{(UV-BC)}  \label{UVBC}
\end{eqnarray}
(hereafter, we consider the chiral limit $m_0=0$).

The general solution of Eq.(\ref{linearized:eq}) has the form
\begin{equation}
B_0(x)/B_0 =C_1u_1(x)+C_2u_2(x),
\end{equation}
where as for two independent solutions of the differential equation
we take
\begin{eqnarray}
 u_1(x) &\equiv& 
 \hyper{\omega(1+i\nu)}{\omega(1-i\nu)}{1+2\omega}
       {-\dfrac{x}{B_0^2}}, \\
 u_2(x) &\equiv& \left(\dfrac{x}{B_0^2}\right)^{-\omega(1+i\nu)}
 \hyper{\omega(1+i\nu)}{-\omega(1-i\nu)}{1+2i\omega\nu}
       {-\dfrac{B_0^2}{x}}+ \mbox{c.\,c.}
\end{eqnarray}
(we consider the case $\kappa_D>\kappa_D^{\rm crit}$).

The boundary conditions (\ref{IRBC}), (\ref{UVBC}) lead to the
following equation determining the mass spectrum:
\begin{equation}
\phi=A_1B_2-A_2B_1=0,
\end{equation}
where we defined the functions $A_i,B_i$ as
\begin{equation}
A_i = \left.\left(1+\frac{1}{2\omega}x\frac{d}{dx}\right)
         u_i(x)\right|_{x=\Lambda^2}, \qquad
B_i = \left.x\dfrac{du_i(x)}{dx}\right|_{x=M_0^2}.
\end{equation}
Using the formulas for differentiating hypergeometric functions
\cite{Bateman}, the functions $A_i,B_i$ can be recasted as
\begin{eqnarray}
&&\hspace{-1cm}A_1=\hyper{\omega(1+i\nu)}{\omega(1-i\nu)}{2\omega}
       {-\dfrac{\Lambda^2}{B_0^2}},\\
&&\hspace{-1cm}A_2={\rm
       Re}\left[\hspace{-1mm}(1-i\nu)\hspace{-1mm}\left(\dfrac{B_0^2}
{\Lambda^2}\right)^{\omega(1+i\nu)}
\hspace{-3mm}\hyper{\omega(1+i\nu)}{1-\omega(1-i\nu)}{1+2i\omega\nu}
       {-\dfrac{B_0^2}{\Lambda^2}}\hspace{-1mm}\right]\hspace{-1mm},
\end{eqnarray}
\begin{eqnarray}
&&\hspace{-1cm}B_1=-\dfrac{\omega^2(1+\nu^2)}{1+2\omega}\dfrac{M_0^2}{B_0^2}\hyper
{1+\omega(1+i\nu)}{1+\omega(1-i\nu)}{2+2\omega}
       {-\dfrac{M_0^2}{B_0^2}},\\
&&\hspace{-1cm}B_2=-2\omega{\rm Re}\left[
  (1+i\nu)\left(\dfrac{B_0^2}{M_0^2}\right)^{\omega(1+i\nu)}
  \hspace{-3mm}
  \hyper{1+\omega(1+i\nu)}{-\omega(1-i\nu)}{1+2i\omega\nu}
        {-\dfrac{B_0^2}{M_0^2}}\right] .
\end{eqnarray}
Since we always assume that $B_0/\Lambda\ll 1$ we can use for $A_i$
their asymptotic expressions
\begin{eqnarray}
A_1&\simeq&|c_0|\sqrt{1+\nu^2}\left(\dfrac{B_0^2}{\Lambda^2}\right)^{\omega}
\sin\left(\omega\nu\ln\dfrac{\Lambda^2}{B_0^2}+\tan^{-1}\nu+\theta\right),\\
\label{A1_asympt}
A_2&\simeq&\sqrt{1+\nu^2}\left(\dfrac{B_0^2}{\Lambda^2}\right)^{\omega}
\cos\left(\omega\nu\ln\dfrac{\Lambda^2}{B_0^2}+\tan^{-1}\nu\right),
\label{A2_asympt}
\end{eqnarray}
where $c_0$ and $\theta$,
\begin{equation}
\theta=\arg\frac{\Gamma(1+2i\omega\nu)}{\Gamma(\omega(1+i\nu))\Gamma(1+\omega(1+i\nu))},
\end{equation}
are defined after Eqs. (\ref{eq:asympt2}), (\ref{eq:sol_bx}).
For the function $B_1$ we use the formula (2.10.2) from \cite{Bateman}
in order to rewrite it in the form similar to $B_2$, thus we have
\begin{eqnarray}
B_1 &=& 2\omega\sqrt{1+\nu^2}|c_0|\left(\dfrac{B_0^2}{M_0^2}\right)^{\omega}
       {\rm Im}\left\{\left(\dfrac{B_0^2}{M_0^2}\right)^{i\omega\nu}
               e^{-i\theta+i\tan^{-1}\nu}\right. \nonumber\\
&& \times\left.
   \hyper{1+\omega(1+i\nu)}{\omega(-1+i\nu)}{1+2i\omega\nu}
         {-\dfrac{B_0^2}{M_0^2}}\right\}, \label{B1}\\
B_2 &=& -2\omega\sqrt{1+\nu^2}\left(\dfrac{B_0^2}{M_0^2}\right)^{\omega}
       {\rm Re}\left\{\left(\dfrac{B_0^2}{M_0^2}\right)^{i\omega\nu}
               e^{i\tan^{-1}\nu}\right. \nonumber\\
&& \times\left.
   \hyper{1+\omega(1+i\nu)}{\omega(-1+i\nu)}{1+2i\omega\nu}
         {-\dfrac{B_0^2}{M_0^2}}\right\}. \label{B2}
\end{eqnarray}
Combining Eqs.(\ref{A1_asympt}),(\ref{A2_asympt}),(\ref{B1}),(\ref{B2}),
the gap equation is transformed to the form
\begin{eqnarray}
\phi &\simeq& -2\omega|c_0|(1+\nu^2)\cos\theta
 \left(\dfrac{M_0^2}{\Lambda^2}\right)^\omega
  \left(\dfrac{B_0^2}{M_0^2}\right)^{2\omega} \nonumber \\
&& \times
  {\rm Im}\left[\left(\dfrac{\Lambda^2}{M_0^2}\right)^{i\omega\nu}
  e^{2i\tan^{-1}\nu}
   \hyper{1+\omega(1+i\nu)}{\omega(-1+i\nu)}{1+2i\omega\nu}
         {-\dfrac{B_0^2}{M_0^2}}\right]=0.
\label{gapeqwithIRcut}
\end{eqnarray}
One can convince oneself that for $M_0\ll B_0$ the last equation is
equivalent to 
\begin{equation}
\sin\left(\omega\nu\ln\dfrac{\Lambda^2}{B_0^2}+\theta+\tan^{-1}\nu\right)=0,
\end{equation}
which gives solutions (\ref{eq:sol_b0}). On the other hand, for
$B_0\ll M_0$ we can use a power expansion of the hypergeometric function
to get the equation for the dynamical mass near the phase transition
point:
\begin{eqnarray}
&&\sin\left(\omega\nu\ln\dfrac{\Lambda^2}{M_0^2}+2\tan^{-1}\nu\right)+
\omega\left[\dfrac{(1+\nu^2)((1+\omega)^2+\omega^2\nu^2)}{1+4\omega^2\nu^2}
\right]^{1/2}\dfrac{B_0^2}{M_0^2}\nonumber\\
&&\times\sin\left(\omega\nu\ln\dfrac{\Lambda^2}{M_0^2}-\tan^{-1}2\omega\nu
+\tan^{-1}\dfrac{\omega\nu}{1+\omega}+\tan^{-1}\nu\right)=0.
\end{eqnarray}
The nontrivial solution for the dynamical mass arises when
$\nu\equiv\sqrt{\kappa_D/\kappa_D^{\rm crit}-1}$ 
exceeds the critical value determined by the
equation
\begin{equation}
\omega\nu_c\ln\dfrac{\Lambda^2}{M_0^2}+2\tan^{-1}\nu_c=\pi.
\end{equation}
It should be noted that $\nu_c$ is a small number
\begin{equation}
\nu_c\simeq\dfrac{\pi}{2+\omega\ln\dfrac{\Lambda^2}{M_0^2}}\ll 1,
\end{equation}
for $\Lambda\gg M_0$. Note also that the form of the gap equation (\ref{gapeqwithIRcut})
is different in two regions $M_0\ll B_0$ and $M_0\gg B_0$: while in the
first one ($M_0\ll B_0$) we observe oscillations in the mass variable,
in the second one ($M_0\gg B_0$) such oscillations disappear. This is
reflected in the character of the mass dependence on the coupling
constant (compare Eqs. (\ref{eq:sol-sd1}) and (\ref{dyn_mass:MF}) below).
In general, it can be shown that Eq. (\ref{gapeqwithIRcut})
has $n$ nontrivial solutions where the number $n$ is given by
\begin{equation}
n=[\pi^{-1}\omega\nu\ln\dfrac{\Lambda^2}{M_0^2}+2\tan^{-1}\nu],
\end{equation}
and the symbol $[C]$ means the integer part of the number $C$.

Near the critical point we thus obtain the mean-field scaling relation
for the dynamical mass
\begin{equation}
\dfrac{B_0^2}{M_0^2}=\dfrac{\pi}{\omega(2\omega^2+2\omega+1)}\dfrac{\nu-\nu_c}
{\nu_c^2}, \quad \nu\gtrsim \nu_c,\quad \nu_c \ll 1,
\label{dyn_mass:MF}
\end{equation}
which is cited in Eq. (\ref{eq:scalingM_0}).

\section{Formulas of the effective gauge coupling strength}
\label{eff_coupl}

The relation between effective and $\overline{\rm MS}$ couplings
($D=4+2$) is
given by Eq.(\ref{eq:eff2}) where the terms $K_g$, $K_b$, $K_f$ denote
one-loop contributions from gauge bosons, gauge scalars, and fermions, 
respectively.
The formulas of $K_g$, $K_b$, and $K_f$ are given by
\begin{eqnarray}
K_g(q^2,\Lambda_g^2)
  &\equiv& 4 C_G \left(
      \frac{5}{18}+\frac{1}{6} \ln \frac{\Lambda_g^2}{(-q^2)}
     +\frac{\Lambda_g^2}{(-q^2)} \tilde K_g(q^2, \Lambda_g^2)
  \right), \\
K_b(q^2,\Lambda_g^2)
  &\equiv&  -2 C_G \left(
      \frac{31}{450}+\frac{1}{30} \ln \frac{\Lambda_g^2}{(-q^2)}
     +\frac{\Lambda_g^2}{(-q^2)}\tilde K_b(q^2, \Lambda_g^2)
  \right), \\
K_f(q^2,\Lambda_g^2)
  &\equiv& - 2\eta T_R N_f \left(
      \frac{47}{900}+\frac{1}{30} \ln \frac{\Lambda_g^2}{(-q^2)}
     +\frac{\Lambda_g^2}{(-q^2)}\tilde K_f(q^2, \Lambda_g^2)
  \right),
\end{eqnarray}
where the sum of the KK-modes is approximated by replacing it to an
corresponding integral.
The functions $\tilde K_i$ ($i=g,b,f$) are defined by
\begin{eqnarray}
  \tilde K_g(q^2, \Lambda_g^2) 
  &\equiv& \int_0^1 dx f(q^2, \Lambda_g^2, x), 
  \nonumber \\
  &=&
   -\frac{4}{3}+\frac{5}{18} \frac{q^2}{\Lambda_g^2}
   +\frac{1}{3}\left(4-\frac{q^2}{\Lambda_g^2}\right)^{3/2}
    \left(\frac{\Lambda_g^2}{-q^2}\right)^{1/2}
    \arctanh \sqrt{\frac{-q^2}{4\Lambda_g^2-q^2}},
  \\
  \tilde K_b(q^2, \Lambda_g^2) 
  &\equiv& \int_0^1 dx (2x-1)^2 f(q^2, \Lambda_g^2, x),
  \nonumber \\
   &=&
   \frac{16}{15}\frac{\Lambda_g^2}{q^2} -\frac{28}{45}
  +\frac{31}{450}\frac{q^2}{\Lambda_g^2}
  +\frac{1}{15}\left(4-\frac{q^2}{\Lambda_g^2}\right)^{5/2}
   \left(\frac{\Lambda_g^2}{-q^2}\right)^{3/2}
   \arctanh \sqrt{\frac{-q^2}{4\Lambda_g^2-q^2}},
   \nonumber\\ 
   & & \\
  \tilde K_f(q^2, \Lambda_g^2) 
  &\equiv& \int_0^1 dx x(1-x) f(q^2, \Lambda_g^2, x), 
  \nonumber \\
  &=&
  -\frac{4}{15}\frac{\Lambda_g^2}{q^2} -\frac{8}{45}
  +\frac{47}{900}\frac{q^2}{\Lambda_g^2}
  \nonumber \\
  & &
  -\frac{1}{15}\left(4-\frac{q^2}{\Lambda_g^2}\right)^{3/2}
   \left(1+\frac{q^2}{\Lambda_g^2}\right)
   \left(\frac{\Lambda_g^2}{-q^2}\right)^{3/2}
   \arctanh \sqrt{\frac{-q^2}{4\Lambda_g^2-q^2}},
\end{eqnarray}
with $f$ being
\begin{displaymath}
 f(q^2,\Lambda_g^2,x) \equiv
   (1-x(1-x)\frac{q^2}{\Lambda_g^2})
   \ln\left(1-\frac{q^2}{\Lambda_g^2}x(1-x)\right).
\end{displaymath}

%
%
\begin{figure*}[t]
  \begin{center}
    \resizebox{0.45\textwidth}{!}{\includegraphics{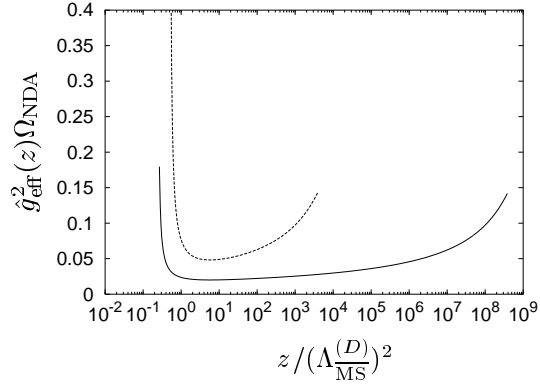}}
    \caption{The effective coupling in the strongly interacting phase
      ($(\Q)^2>0$). 
             The solid and dashed lines represent the graphs for 
             $(\Lambda_g R)^2=4 \times 10^{10}, 4 \times 10^5$ 
             with $N_f=2$, $N=3$, $(\Q)^2=10^2$, respectively. 
             \label{fig:geff_Q}}
  \end{center}
\end{figure*}

We next discuss behavior of the effective coupling 
$\hat g_{\rm eff}^2(z)$ in the energy region $z \ll \Lambda^2$.
Expanding the function $f$ around $q^2=0$
\begin{displaymath}
  f(q^2,\Lambda_g^2,x) = -x(1-x)\dfrac{q^2}{\Lambda_g^2}
                         +\frac{1}{2} x^2(1-x)^2 \dfrac{q^4}{\Lambda_g^4}
                         +{\cal O}\left(
                           \left(\dfrac{q^2}{\Lambda_g^2}\right)^3
                          \right),
\end{displaymath}
we find 
\begin{eqnarray*}
 \tilde{K}_g(q^2, \Lambda_g^2)
&=& -\frac{1}{6}\frac{q^2}{\Lambda_g^2} 
    + \frac{1}{60}\left(\frac{q^2}{\Lambda_g^2}\right)^2 
    + {\cal O}\left(\left(\frac{q^2}{\Lambda_g^2}\right)^3\right), \\
 \tilde{K}_b(q^2, \Lambda_g^2)
&=& -\frac{1}{30}\frac{q^2}{\Lambda_g^2} 
    + \frac{1}{420}\left(\frac{q^2}{\Lambda_g^2}\right)^2 
    + {\cal O}\left(\left(\frac{q^2}{\Lambda_g^2}\right)^3\right), \\
 \tilde{K}_f(q^2, \Lambda_g^2)
&=& -\frac{1}{30}\frac{q^2}{\Lambda_g^2} 
    + \frac{1}{280}\left(\frac{q^2}{\Lambda_g^2}\right)^2
    + {\cal O}\left(\left(\frac{q^2}{\Lambda_g^2}\right)^3\right).
\end{eqnarray*}
It is then easy to obtain
\begin{equation}
  \sum_{i=g,b,f} K_i(-z, \Lambda_g^2)
  = \left( \frac{3}{5} C_G -\frac{\eta}{15} T_R N_f \right)
    \ln\dfrac{\Lambda_g^2}{z}
   + \dfrac{1}{75} \left(118 C_G -\frac{77}{6} \eta T_R N_f\right)
   +  {\cal O}\left(\frac{z}{\Lambda_g^2}\right).
  \label{eq:sum_K}
\end{equation}
Plugging Eq.(\ref{eq:sum_K}) into Eq.(\ref{eq:g_eff_K}), we find
\begin{equation}
  \dfrac{1}{\hat g_{\rm eff}^2(z)}
  = -\dfrac{1}{g_*^2} \dfrac{(\Q)^2}{z} 
    +\dfrac{1}{(4\pi)^3}\left( \frac{3}{5} C_G -\frac{\eta}{15} T_R N_f \right)
    \ln\dfrac{\Lambda_g^2}{z}
    +\cdots.
\end{equation}
\end{widetext} 
Note here that the effective coupling $\hat g_{\rm eff}^2$ depends on
the ultraviolet cutoff $\Lambda_g$, indicating the violation of the
decoupling theorem.

It is now easy to see that the effective coupling $\hat g_{\rm eff}^2$ 
remains finite in the infrared region $0 \leq z \ll \Lambda_g^2$ for $(\Q)^2<0$
for $C_G > \eta T_R N_f/9$, which is automatically satisfied when
$-b'=(10C_G - \eta T_R N_f)/3 >0$. 
On the other hand, Fig.~\ref{fig:geff_Q} shows typical behaviors of
the effective coupling with $(\Q)^2>0$.
We find that the effective coupling diverges at $z \sim (\Q)^2 >0$ in
this case.

\end{document}